\documentclass[10pt,letterpaper]{article}
\usepackage[]{geometry}

\usepackage{amsmath,amssymb}

\usepackage{changepage}

\usepackage[utf8x]{inputenc}
\usepackage[T1]{fontenc}
\usepackage{lmodern}

\usepackage{textcomp,marvosym}

\usepackage{cite}

\usepackage{nameref}
\usepackage[hidelinks]{hyperref}

\usepackage[right]{lineno}

\usepackage{microtype}
\DisableLigatures[f]{encoding = *, family = * }

\usepackage[table]{xcolor}

\usepackage{array}
\usepackage{float} 

\usepackage{csquotes}
\usepackage{siunitx}
\usepackage[version=4]{mhchem}

\newcolumntype{+}{!{\vrule width 2pt}}

\newlength\savedwidth

\usepackage[aboveskip=1pt,labelfont=bf,labelsep=period,justification=raggedright,singlelinecheck=off]{caption}

\makeatletter
\renewcommand{\@biblabel}[1]{\quad#1.}
\makeatother

\usepackage{lastpage,fancyhdr,graphicx}
\usepackage{epstopdf}
\fancyheadoffset[L]{2.25in}
\fancyfootoffset[L]{2.25in}
\DeclareMathOperator*{\argmax}{arg\,max}

\begin{document}
\vspace*{0.2in}

\begin{flushleft}
{\Large
\textbf\newline{Information-theoretic analyses of neural data to minimize the effect of researchers' assumptions in predictive coding studies} 
}
\newline
\\
Patricia Wollstadt\textsuperscript{1}
Daniel L. Rathbun\textsuperscript{2,3},
W. Martin Usrey\textsuperscript{2,4},
Andr\'e Moraes Bastos\textsuperscript{5},
Michael Lindner\textsuperscript{7},
Viola Priesemann\textsuperscript{6},
Michael Wibral\textsuperscript{7}
\\
\bigskip
\textbf{1} MEG Unit, Brain Imaging Center, Goethe University, Frankfurt/Main, Germany
\\
\textbf{2} Center for Neuroscience, University of California, Davis, CA, United States 
\\
\textbf{3} Center for Ophthalmology, University of T\"uebingen, T\"uebingen, Germany
\\
\textbf{4} Department of Neurobiology, Physiology, and Behavior, University of California, Davis, CA, United States
\\
\textbf{5} Department of Psychology and Vanderbilt Brain Institute, Vanderbilt University, Nashville, TN, United States
\\
\textbf{6}
Max Planck Institute for Dynamics and Self-Organization, G\"ottingen, Germany
\\
\textbf{7}
Campus Institute for Dynamics of Biological Networks, University G\"ottingen, G\"ottingen, Germany
\\
\bigskip

%
%





* patricia.wollstadt@gmx.de
\end{flushleft}

\section*{Abstract}
Studies investigating neural information processing often implicitly ask both, which processing strategy out of several alternatives is used and how this strategy is implemented in neural dynamics. A prime example are studies on predictive coding. These often ask whether confirmed predictions about inputs or predictions errors between internal predictions and inputs are passed on in a hierarchical neural system---while at the same time looking for the neural correlates of coding for errors and predictions. If we do not know exactly what a neural system predicts at any given moment, this results in a circular analysis---as has been criticized correctly. To circumvent such circular analysis, we propose to express information processing strategies (such as predictive coding) by local information-theoretic quantities, such that they can be estimated directly from neural data. We demonstrate our approach by investigating two opposing accounts of predictive coding-like processing strategies, where we quantify the building blocks of predictive coding, namely predictability of inputs and transfer of information, by local active information storage and local transfer entropy. We define testable hypotheses on the relationship of both quantities, allowing us to identify which of the assumed strategies was used. We demonstrate our approach on spiking data collected from the retinogeniculate synapse of the cat ($N=16$). Applying our local information dynamics framework, we are able to show that the synapse codes for predictable rather than surprising input. To support our findings, we estimate quantities applied in the partial information decomposition framework, which allow to differentiate whether the transferred information is primarily bottom-up sensory input or information transferred conditionally on the current state of the synapse. Supporting our local information-theoretic results, we find that the synapse preferentially transfers bottom-up information.

\section*{Introduction}

Predictive coding as a theory arguably dominates today's scientific discourse on how the cortex works \cite{Clark2013, Hohwy2013book, Hawkins2007}. Importantly, it is positioned as a theory of general cortical function---yet, empirical tests so far are limited to situations with an explicitly predictive experimental context, simply to allow for a meaningful analysis. In other words, to find and understand the neurophysiological correlates of predictions and errors, experiments posit a priori, when and what is being predicted in which brain region. There are three problems with this approach: first, knowing what is being predicted when and where in the brain seems to require already a fair understanding of how the brain, or the cortex, works---which may not generally be available yet. Second, trying to acquire some of the necessary knowledge post-hoc, runs the real risk of involuntarily producing a circular analysis or argument, or a \enquote{just-so story} (as it is called e.g. in section 4.1 of \cite{litwin2020unification}). Third, restricting empirical tests of a general theory to experimental contexts that are explicitly designed with predictions in mind, in a strict sense, prohibits conclusions about the applicability of that theory in other contexts. One might provocatively frame this third problems as: \enquote{Is the cortex doing predictive coding when we don't test it?} Last but not least, the latter restriction to dedicated experimental designs excludes testing (and possibly refuting) the theory by drawing on the vast majority of empirical neurophysiological data, i.e., all data that were obtained with a focus on descriptions of cortical function(s) other than predictive coding, which seems like a waste of available empirical evidence.

In this paper, we introduce an information-theoretic framework for testing predictive coding theories by translating the concepts of predictability, predictions, and prediction errors (surprise) into information-theoretic quantities measurable from data. Based on these information-theoretic formulations, we describe how to test by pure information-theoretical means whether a neural processing element (neuron, or a small circuit) takes part in a predictive coding-like computation. Our novel method is in principle applicable without knowledge of the intentions of the experimenter, given some weak constraints on the data themselves. Our method rests on the simple idea that a neural processing element that codes for prediction errors should exhibit high information transfer at moments when its input is surprising (i.e. fundamentally unpredictable), and vice versa \cite{Wibral2015BitsFromBrains}. On the other hand, a processing element coding for the predictable information in its inputs should exhibit high information transfer at times when this predictable information is high.

In the following we formalize this idea in the mathematical context of local information dynamics \cite{lizier2013book} and partial information decomposition \cite{Williams2010}.

\section*{Methods}

\subsection*{Local information dynamics}
As local information dynamics \cite{lizier2013book} is a relatively recent subfield of information theory, such that the inspection of local information quantities is not yet widely applied, we (re-)introduce these concepts with some detail. In this exposition we try to balance a concise and intuitive presentation of the material with mathematical rigor, necessarily sacrificing some of the latter. For a more detailed introduction see \cite{Wibral2015BitsFromBrains,lizier2013book}.

\subsection*{Local mutual information}
For the purpose of this study it is best to understand the mutual information, $I(X:Y)$, between two random variables by stating that if one variable $X$ has information about another variable, $Y$, then $X$ and $Y$ can not be statistically independent. This point of view will help to understand why each individual term, $\log\frac{p(x,y)}{p(x)p(y)}$, that contributes to the summation in the mutual information:

\begin{equation}
	I(X:Y) = \sum_{x,y \in \mathcal{A}_{X,Y}}p(x,y) \log\frac{p(x,y)}{p(x)p(y)} ~,
	\label{eq:MI}
\end{equation}

\noindent can be soundly interpreted on its own. By $x \in \mathcal{A}_{X}$ and $y \in \mathcal{A}_{Y}$, we denote individual realizations of random variables $X$ and $Y$, by and we write $p(x)$ as a shorthand for the probability $p(x=X)$. We start our explanation with the definition of statistical independence of $X$ and $Y$ as:

\begin{equation}
	p(x,y) = p(x)p(y) \quad \forall x,y \in \mathcal{A}_{X,Y}~,
	\label{eq:indep}
\end{equation}

\noindent meaning that the equation $p(x,y) = p(x)p(y)$ must hold for \emph{all} pairs of realizations $(x,y)$. If the above equation is violated for any pair $(x,y)$ then this pair contributes to a deviation from independence. As per our initial statement on the relation of independence and information, this pair then also contributes to the information that $X$ holds about $Y$, and vice versa.

To measure how much independence is violated locally by the pair $(x,y)$, we can take the ratio of both sides of Eq~\ref{eq:indep}. Now, independence, or the absence of mutual information, is equivalent to:

\begin{equation}
	\frac{p(x,y)}{p(x)p(y)}=1 \quad \forall x,y \in \mathcal{A}_{X,Y}~.
	\label{eq:ratio}
\end{equation}

\noindent A deviation of this ratio from $1$ for any pair $(x,y)$ indicates a deviation from independence, i.e., the presence of information in the realization of one variable about the realization of the other. Obviously one would like a measure of this information itself to be zero in the absence of information, i.e., at independence. This can be achieved by taking the logarithm of Eq~\ref{eq:ratio}:

\begin{equation}
	\log\frac{p(x,y)}{p(x)p(y)}=0 \quad \forall x,y \in \mathcal{A}_{X,Y}~.
	\label{eq:measure}
\end{equation}

\noindent Inspecting equation \ref{eq:measure} and comparing to the definition of the mutual information in equation \ref{eq:MI}, we now see that the mutual information is nothing but the weighted average of the individual deviations from independence, measured on a logarithmic scale. More importantly the above derivation of the mutual information demonstrates that each individual term has a well defined and interpretable meaning. These individual terms define the local mutual information in a pair of realizations, $i(x,y)$:

\begin{equation}
i(x:y)=\log\frac{p(x,y)}{p(x)p(y)}=\log\frac{p(x|y)}{p(x)}~.
\end{equation}

\noindent Analogously, the local conditional mutual information between two variables in the context of a third is given by:

\begin{equation}
i(x:y|z)=\log\frac{p(x,y|z)}{p(x|z)p(y|z)}=\log\frac{p(x|y,z)}{p(x|z)}~.
\end{equation}

We note that the local interpretation introduced here is closely related to the way Fano originally derived the mutual information \cite{Fano1961}. In addition, we note that the local mutual information and the local conditional mutual information can be negative---in contrast to the (average) mutual information which is always positive or zero. We will explain this fact in detail, and make use of it, further below.

\subsection*{Local active information storage as locally predictable information}

\begin{figure}[!h]
	\includegraphics{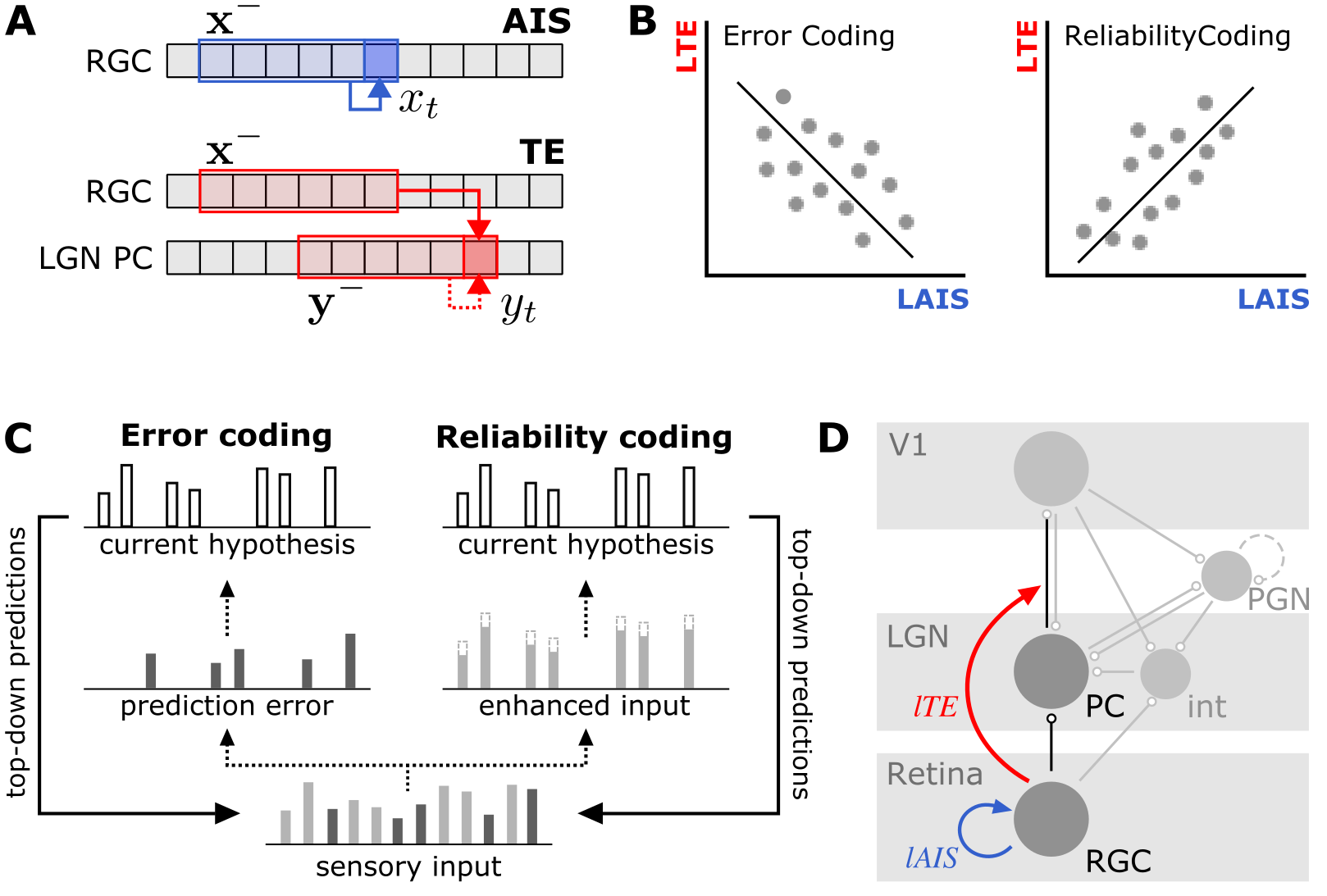}  
	\caption{
		{\bf Overview of analysis approach.}
	    \textbf{A} Information-theoretic measures of predictability and information transfer: active information storage ($AIS$) quantifies the predictability of a processes' current state $x_t$ from its immediate past $\mathbf{x}^-$, transfer entropy (TE) quantifies the information transfer from a source process $X$ to a target process $Y$ by quantifying the predictability of the target's current state, $y_t$ from the sources' past, $\mathbf{x}^-$, in the context of the target's immediate past, $\mathbf{y}^-$.
		\textbf{B} Local storage-transfer correlations (LSTC) relating local AIS ($lAIS$) as a measures of predictability and local TE ($lTE$) as a measure of information transfer: if a neuron codes for predictable input a positive correlation is expected, if the neuron codes for unpredictable input, a negative correlation is expected (adapted from \protect\cite{Wibral2015BitsFromBrains}).
		\textbf{C} Realizations of predictive coding in the cortex (adapted from \protect\cite{Murray2004}): bottom-up sensory input (dotted arrows) is compared to predictions propagated in top-down direction from a hierarchically higher cortical level (solid arrows) that represent the current prior about the input (white bars). See main text.
		\textbf{D} Physiology of the retinogeniculate synapse and recording sites \protect\cite{Ahlsen1985,Rogala2013}:
		Recordings were collected from in- and outputs to the synapse between retinal ganglion cells (RGC) and Layer A principal cells (PC) in the lateral geniculate nucleus (LGN). We estimated local active information storage ($lAIS$, blue arrow) within the synapse input, and local transfer entropy ($lTE$, red arrow) between in- and output of the synapse. Schematic representations of known connections of PC and RGC are shown in grey (round markers indicate synapses): excitatory cells in layer 6 of primary visual cortex (V1) form feedback connections with LGN PC and also project to LGN inhibitory interneurons (int) and perigeniculate nucleus (PGN). Interneurons provide inhibitory input to LGN PC: intrageniculate interneurones (int) mediate feed-forward inhibition from RGC cells, while PGN cells provide recurrent inhibition \cite{Dubin1977,Ahlsen1985}; PGN interneurons further form reciprocal, inhibitory connections amongst each other (dashed line).
	}
	\label{fig:introduction}
  \end{figure}

Using the local mutual information defined above, we can quantify how much of the information contained in a process (e.g., a neural signal) at the present moment $t$ is predictable from its past. We assume that such a process denotes an ordered collection of random variables, $\mathbf{X} = \left\{X_t\right\}$, with realizations $x_t\in\mathcal{A}_{X_t}$. We then quantify how predictable a single realization, $x_t$, is from its past in the following way:

\begin{equation}
    \begin{split}
    lAIS(x_t : \mathbf{x^{-}})\equiv & i(x_t:\mathbf{x^{-}}) \\
    =&\log\frac{p(x_t|\mathbf{x^{-}})}{p(x_t)}~,
    \end{split}
    \label{eq:lais}
\end{equation}

\noindent where $lAIS$ is shorthand for the local active information storage \cite{lizier2012lais}, and $\mathbf{x^{-}}$ is a realization of the (possibly infinite) past of the process up to time $t$ (Fig~\ref{fig:introduction}, \textbf{A}).

As already mentioned above the local mutual information forming the local active information storage need not be positive, i.e.\,:

\begin{equation}
	lAIS(x_t: \mathbf{x^{-}})<0 ~\Leftrightarrow~ p(x_t|\mathbf{x^{-}})<p(x_t).
\end{equation}

\noindent This means that a negative $lAIS$ indicates that the $x_t$ that actually happened was less expected to happen, given the information in the past of the process, $\mathbf{x^{-}}$, than it was expected to happen without this information about the past. Put differently, the past $\mathbf{x^{-}}$ \emph{mispredicted} the actual value of $x_t$ by allocating the probability mass originally contained in $p(x_t)$ to other values in the conditional distribution. Since we assume that all probabilities, including the conditional probabilities above, are computed properly, this is a necessary misprediction, not one that could have been avoided. In other words, negative $lAIS$ indicates unpredictable behavior of the process at time $t$. If we think of the overall process $\mathbf{X}$ as the input to a neuron then negative $lAIS$ at time $t$ means that the neuron must mispredict at $t$, based on the past of its inputs.

As $lAIS$ is a relatively novel concept in the analysis of neural information processing, a few words on its interpretation and validity in a biological context are necessary. In particular, it is important to consider potential sources of predictability and in what respect this origin of predictability matters. There are in general two possible sources of predictability in a neural spike train: first, predictability arising from temporal statistical dependencies in the input, and second, internally generated temporal dependencies. Does one of these matter more than the other? To answer this question it makes sense to take the point of view of the neuron receiving the spike train. This neuron, throughout its existence has received nothing but the incoming spikes, and has no access to any \enquote{ground-truth} about the outside world, and temporal regularities in this outside world. So from a neural information processing perspective, the above distinction must vanish. Thus for the analysis framework presented here, it does in principle not matter whether the stimulus input to the retina was predictable or not; all that matters is whether the incoming spike train from the retina was predictable at the level of the LGN. So from a neuro-centric perspective an analysis without the stimulus properties seems to us to reflect the circumstances a neuron finds itself in.

As we will also see below the predictability mainly resided in the inter spike interval (ISI) distribution of the incoming RGC spikes. However, this is not problematic as predictability has to reside somewhere in the physical and statistical properties of the spike train. This is because in some sense information theory is \enquote{only} a summary statistic of some probability distribution (here, the joint distribution of past and present spiking activity). Yet, it is a very special summary statistic in the sense that it truly quantifies the relevant amount of information that is stored in or transmitted by a signal---and the respective information-theoretic measures are the only consistent measures of this information. In this sense, information theory simply puts an \enquote{information processing price tag} on some biophysical process. In other words, methods do not learn anything not already contained in the respective probability distributions, but using information theory the information processing hidden in these distributions becomes visible and quantifiable.

After defining predictability formally via $lAIS$ and reasoning for the applicability of this new concept,  all that is left to measure now is whether the neuron is transferring mispredicted information onwards (as in coding for prediction errors) or whether it does not transfer information at those moments when the input is mispredicted (as in coding for the predictable input information only). A measurement of how much information is transferred by a neuron at each moment in time is given by the local transfer entropy \cite{lizier2008lte}.

\subsection*{Measuring transmitted information as local transfer entropy from inputs to output}

The information flowing from input process(es) (e.g., inputs to a neuron), $\mathbf{X}$, to an output process, $\mathbf{Y}$, (e.g., output of a neuron) at any moment in time, $t$, is given by the local transfer entropy \cite{lizier2008lte} (Fig~\ref{fig:introduction}, \textbf{A}):

\begin{equation}
    \begin{split}
	lTE(\mathbf{x^{-}}\rightarrow y_t) &\equiv i(y_t:\mathbf{x^{-}}|\mathbf{y^{-}}) \\
	& =\log\frac{p(y_t|\mathbf{x^{-}},\mathbf{y^{-}})} {p(y_t|\mathbf{y^{-}})}~,
	\end{split}
	\label{eq:lte}
\end{equation}

\noindent which quantifies the information transferred from the input's past state, $\mathbf{x^{-}}$, about the present state in $Y$, $y_t$, in the context of $\mathbf{Y}$'s immediate past state, $\mathbf{y^{-}}$. Again, the $lTE$ can be negative; in this case the negativity indicates that there is information in the output that is unexpected, given the past input from $\mathbf{X}$,

\begin{equation}
	lTE(\mathbf{x^{-}}\rightarrow y_t) < 0 ~\Leftrightarrow~ p(y_t|\mathbf{x^{-}},\mathbf{y^{-}}) < p(y_t|\mathbf{y^{-}}),
\end{equation}

\noindent i.e., negative $lTE$ indicates that the $y_t$ that actually happened was less expected to happen, given both the information in its own immediate past $\mathbf{y}^-$ and the past of the process $\mathbf{x^{-}}$ than without the information in $\mathbf{x^{-}}$. In other words, the past $\mathbf{x}^-$ mispredicted the actual value of $y_t$ given the information obtained from $\mathbf{y}_t$. Similar to $lAIS$, this misprediction quantifies behavior of the process $\mathbf{Y}$ at time $t$ that is unpredictable from the sources' past, in the context of the target's own past.

Using both measures, $lAIS$ and $lTE$, we are now able to quantify locally how predictable the current state of a single process is from its own past. Secondly, we are able to quantify locally, how much information is transferred from one process to a another. Again, we note that the information-theoretic quantities in question must necessarily be driven by stimulus properties and biophysics. Thus, the information-theoretic measures do not tell us something entirely removed from other descriptions of neural activity. Yet, only their use allows for a rigorously quantitative interpretation of the biophysical observations in terms of neural information processing. In other words, a rigorous \emph{quantitative} analysis of neural information processing necessitates the use of proper information-theoretic measures.

Next, we detail how relating both measures enables us to address the important issue of predictive coding.

\subsection*{Local storage-transfer correlations (LSTCs) as an indicator for predictive processing}

Using both $lAIS$ on the input to a neuron and the $lTE$ between its input and its output, we can now assess whether a neuron performs a predictive-coding like computation \cite{lizier2012coherent,Wibral2015BitsFromBrains} (Fig~\ref{fig:introduction}, \textbf{B} and \textbf{C}). More specifically:

\begin{itemize}
	\item If a neuron codes for the predictable parts of its input it should have a \emph{highly positive} local information transfer $lTE$ from input to output at moments $t$ when the predictability of the input as measured by $lAIS$ is \emph{high}. That is, the correlation between $lTE$ and $lAIS$ should be \emph{positive}.
	\item If, in contrast, a neuron codes for the unpredictable parts of its input, then local information transfer $lTE$ should be 
	\emph{highly positive} at moments $t$ when the  predictability of the input as measured by $lAIS$ is \emph{very low} or even \emph{negative}. That is, the correlation between $lTE$ and $lAIS$ should be \emph{negative}.
\end{itemize}

\noindent The first variant of predictive coding theory has been proposed, for example, in adaptive  resonance  theory  (ART) \cite{Grossberg1980,Grossberg2013} or  the  biased competition model \cite{Desimone1995,Desimone1998}, which assume that the signaling of bottom-up sensory evidence is facilitated for sensory input that matches the current predictive model (Fig~\ref{fig:introduction}, \textbf{C}). The second variant has been proposed, for example, in  \cite{Rao1999,Mumford1992,Hohwy2008}, where it is suggested that bottom-up signals represent prediction errors, i.e., sensory input that is not predicted by the current internal model (Fig~\ref{fig:introduction}, \textbf{C}). Both variants have been shown to be functionally equivalent such that an implementation of predictive coding can be achieved by both \cite{Spratling2008} (see also \cite{Kveraga2007,Siegel2000,Summerfield2009}, and the \textit{Discussion} section).

We want to highlight that the assessment of predictive coding using the above described local storage-transfer correlation (LSTC, Fig~\ref{fig:introduction}, \textbf{B}) requires only minimal knowledge on the experiment that provided the data, namely, it is sufficient to know how to properly assess the probability distributions involved in the estimation of $lAIS$ and $lTE$. The approach is thus applicable to data from a vast range of experiments, including those not specifically designed with predictive coding in mind. Most importantly, this approach does not require knowledge on what the brain or a neuron should predict.

\subsection*{Partial information decomposition as a measure of state-dependent and -independent information transfer}

Before we go on to describe how to estimate LSTC from data, we want to introduce the framework of partial information decomposition (PID) \cite{Williams2010,Williams2011,Gutknecht2020,Makkeh2020,schick2021partial} as a second tool to investigate predictive coding in neural processing. PID is a recent extension to classical information theory. Amongst other applications in neuroscience (e.g., \cite{Timme2014,Timme2016,Wibral2015BitsFromBrains,Wibral2017pid}), PID allows to decompose information transfer measured by transfer entropy (TE) into contributions that are reflective of the calculation of a \enquote{generalized prediction error} versus contributions that indicate a relaying of predictable information.

PID describes how two or more source variables provide information about a target variable, where each source may provide unique information (information that is only available from this particular source), redundant information (information that is available redundantly from two or more sources), and synergistic information (information that is only available when considering two or more input variables together) \cite{Williams2010}. Note that in this study, we apply a non-localized measure of PID \cite{bertschinger2014unq} (discussed in detail below), and therefore refer to averaged quantities only (for first proposals of localized PID measures see\cite{Finn2018,Makkeh2020}).

To illustrate how PID can be used to decompose TE \cite{Williams2011}, we first take a closer look at the calculation of the TE as the conditional mutual information $I(Y_t:\mathbf{X^-}|\mathbf{Y^-})$. Here, conditioning on the target's past state, $\mathbf{Y^{-}}$, influences the information the inputs' past state, $\mathbf{X^{-}}$, provides about $Y_t$ in one of the following manners,

\begin{itemize}
    \item in the context $\mathbf{Y^{-}}$, $\mathbf{X^{-}}$ may provide \emph{less} information about $Y_t$ such that $I(Y_t:\mathbf{X^{-}}|\mathbf{Y^{-}})<I(Y_t:\mathbf{X^{-}})$;
    \item in the context $\mathbf{Y^{-}}$, $\mathbf{X^{-}}$ may provide \emph{more} information about $Y_t$ such that $I(Y_t:\mathbf{X^{-}}|\mathbf{Y^{-}})>I(Y_t:\mathbf{X^{-}})$;
    \item there may be no change in the information provided by $\mathbf{X^{-}}$ about $Y_t$, such that $I(Y_t:\mathbf{X^{-}}|\mathbf{Y^{-}}) = I(Y_t:\mathbf{X^{-}})$, i.e., $Y$'s past is independent of $\mathbf{X^{-}}$ and $Y_t$, and knowing $Y$'s past does not influence the information we obtain from $\mathbf{X^{-}}$ about $Y_t$.
\end{itemize}

These changes in information contribution may be decomposed and quantified using PID terms \cite{Williams2011}: The first case may be interpreted as scenarios in which information about $Y_t$ is redundantly present in both past states, $\mathbf{X^{-}}$ and $\mathbf{Y}^-$, such that by conditioning on $\mathbf{Y}^-$ this redundant information is \enquote{removed} from the information $\mathbf{X^{-}}$ provides about $Y_t$. The second case describes scenarios in which both past states provide synergistic information about $Y_t$, which is \enquote{added} to the information $\mathbf{X^{-}}$ provides uniquely about $Y_t$. Note that both redundant and synergistic information contribution can be simultaneously present in the interaction of two variables with respect to a third. The third case can be loosely thought of as the information $\mathbf{X^{-}}$ entering ${Y_t}$ being both, independent of $\mathbf{Y^{-}}$ and being encoded into  ${Y_t}$ independently of $\mathbf{Y^{-}}$---thus, it reflects a unique information transfer from  $\mathbf{X^{-}}$ to  ${Y_t}$. In sum, when calculating TE, i) we remove redundant information in $\mathbf{X^{-}}$ and $\mathbf{Y}^-$ about $Y_t$, ii) we measure \textit{synergistic} information jointly present in $\mathbf{X^{-}}$ and $\mathbf{Y}^-$ about $Y_t$ \cite{Williams2011}, and iii) we measure the information provided \textit{uniquely} by $\mathbf{X}^-$ about $Y_t$.

\subsection*{State-dependent transfer entropy as generalized prediction error}

PID allows us to decompose information transfer from a cell's inputs, $\mathbf{X}^-$ to its output $Y_t$, conditional on the target's past, $\mathbf{Y}^-$, into different contributions: We can quantify the information uniquely provided by  $\mathbf{X}^-$ about $Y_t$, independently of $\mathbf{Y}^-$, also termed \emph{state-independent} TE \cite{Williams2011}, and we can quantify the information provided by $\mathbf{X}^-$ about $Y_t$ synergistically with $\mathbf{Y}^-$, i.e., dependent on the state of $\mathbf{Y}^-$, also termed \emph{state-dependent} TE \cite{Williams2011}. One may think about the latter case as the target's past state \enquote{decoding} the information transferred from the source to the target.

The computation of state-dependent TE, i.e. the synergistic part of the TE, is of particular relevance here, as the synergy reflects the computation of a \enquote{generalized prediction error} from the past state of the target cell (the prediction) and the past state of the input (the sensory evidence) and the error's transfer by the target neuron. This can best be seen by considering that the computation of a binary error (e.g., in a spiking neuron) is analog to the XOR operation and that this operation leads to synergistic information: here, knowing only one input is not sufficient to know what the output of the system should be---this is only possible if both inputs are considered at once.

\subsection*{Partial information decomposition measures and estimation}

The PID framework \cite{Williams2010} extends classical information theory to allow for the decomposition described in the last section by proposing a set of new axioms. However, the exact choice of PID axioms, which allows the definition of appropriate \textit{measures}, is still an active area of research at the time of writing \cite{griffith2014,harder2013,bertschinger2014unq,Makkeh2020,Gutknecht2020,schick2021partial}. To estimate PID, we here follow a proposal by Bertschinger et al. \cite{bertschinger2014unq} that is based on the estimation of unique information, which is grounded on the assumption that in a suitably chosen decision problem, exploiting the unique information should yield a measurable advantage.

An implementation for estimating Bertschinger et al.'s measure \cite{bertschinger2014unq} has been proposed in \cite{makkeh2018broja,makkeh2017optimization} and is available as part of the IDTxl toolbox \cite{wollstadt2019idtxl}.

\subsection*{Estimation of information-theoretic quantities from data}

Typically, in experimental neuroscience the probability distributions underlying observed data, which are necessary to calculate the quantities introduced above, are unknown and have to be estimated from data. We will therefore introduce the estimation of local information-theoretic quantities and PID terms from data in this section.

The most straightforward approach to estimating (conditional) mutual information from discrete data (Eq~\ref{eq:lais} and Eq~\ref{eq:lte}) is by replacing probability mass functions by the relative frequencies of symbols observed in the data \cite{hlavavckova2007estimators}. These so-called \enquote{plug-in estimators} are well-known to exhibit negative bias for finite data, for which analytic bias-correction procedures exist \cite{miller1955,panzeri1996correction}. These bias-correction approaches, formulated for non-local variants of mutual information, may be adapted for the use with localized measures to obtain locally bias-corrected estimators of $lAIS$ and $lTE$ (see supporting information~S1). Furthermore, statistical testing against estimates from surrogate data may be applied to handle estimator bias \cite{lindner2011,vicente2011} by treating the estimate as a test statistic compared against a Null-distribution generated from estimates from surrogate data.

Before applying estimators, past states of the time series involved have to be defined. In theory, both AIS and TE quantify the information contained in the semi-infinite past of a time series up to, but excluding, time point $t$. In practice, few observed systems actually retain information for an infinitely long time, such that most information is contained within the immediate past of the present system state, $X_t$ \cite{takens1981,ragwitz2002}. Hence, we can define an \enquote{embedding} of the time series, $\mathbf{X}^S$, i.e., a collection of past variables up to a maximum lag, selected such that the embedding is maximally informative about $X_t$. In mathematical terms, we define the embedding, $\mathbf{X}^S$, such that the Markov property

\begin{equation}
	p(X_t|\mathbf{X}^S) \approx p(X_t|X_{t-1}, \ldots, X_{0}),
	\label{eq:markov_condition}
\end{equation}

\noindent is fulfilled for all $X_t$. In other words, $X_t$ becomes conditionally independent of all variables prior to $\mathbf{X}^S$.

Several approaches for defining such an embedding exist. We here propose the use of a \textit{nonuniform} embedding \cite{faes2011nonuniform,lizier2012multivariate}, that selects variables from a set of past candidate variables, $\mathbf{C}$, such that $\mathbf{X}^S$ becomes maximally informative about $X_t$. A suitable algorithm that handles the computational complexity of selecting this set of variables is a greedy forward-selection strategy that maximizes the information contained in the variable set with respect to $X_t$ using the CMI as selection criterion,

\begin{equation}
	C^* = \argmax_C I(X_{t}: C | \mathbf{X}_i^S) \, \forall C \in \mathbf{C}_i,
	\label{eq:max_candidate}
\end{equation}

\noindent where $\mathbf{X}_i^S$ is the set of variables already selected in the $i$th step of the algorithm and $C$ are candidate variables from the set of candidates $\mathbf{C}_i$. The candidate set is defined as a collection of past variables up to time point $t$, $\mathbf{C}=\left\{X_{t-l}, X_{t-l-1}, \ldots, X_{t-k} \right\}$, where $l$ $k$ denote a maximum and minimum lag with respect to $t$. Surrogate testing is used to evaluate whether the selected variable, $C^*$ provides additional information about $X_t$ by testing whether $I(X_{t}: C^* | \mathbf{X}_i^S)$ is statistically significant. If so, $C^*$ is included into the embedding and removed from the set of candidates,

\begin{align}
	\mathbf{C}_{i+1} &\leftarrow \mathbf{C}_i \setminus C^*, \\
	\mathbf{X}^S_{i+1} &\leftarrow \mathbf{X}^S_i \cup C^*.
	\label{eq:set_update}
\end{align}

\noindent For a detailed account of the estimation procedure including a hierarchical statistical testing scheme that handles the family-wise error rate of the repeated testing during the iterative candidate selection, see \cite{novelli2019} and the implementation in \cite{wollstadt2019idtxl}.

The greedy strategy for constructing past states can be directly applied to find a non-uniform embedding of the past of a process $X$, such that we can estimate $lAIS$ as

\begin{equation}
	lAIS(x_t : \mathbf{x^{-}}) = i(x_{t}: \mathbf{x}^S).
	\label{eq:lais_est}
\end{equation}

For the construction of past states for $lTE$ estimation, we first optimize the target embedding, $\mathbf{y}^S$ (which amounts to quantifying the active information storage in the target) \cite{wibral2013timing}, before optimizing the source's embedding in the context of the target embedding,

\begin{equation}
	C^* = \argmax_C I(Y_{t}: C | \left\{\mathbf{Y}^S, \mathbf{X}_i^S \right\}) \, \forall C \in \mathbf{C}_i,
	\label{eq:max_candidate_source}
\end{equation}

\noindent where sets $\mathbf{C}_i$ and $\mathbf{x}_i^S$ are updated according to Eq~\ref{eq:set_update}. We can then estimate $lTE$ as

\begin{equation}
	lTE(\mathbf{x^{-}}\rightarrow y_t) = i(y_{t}: \mathbf{x}^S| \mathbf{y}^S).
	\label{eq:lte_est}
\end{equation}

By first optimizing the target's past state, we make sure that we account for all information $\mathbf{Y}$'s past provides about the current state $Y_t$, before quantifying additional or novel information $\mathbf{X}$ provides about $\mathbf{Y}$. This means that only information actually transferred between $\mathbf{X}$ and $\mathbf{Y}$ is taken into consideration when estimating $lTE$.

We used a software implementations of the proposed approach provided by the IDTxl Python toolbox \cite{wollstadt2019idtxl}, which internally makes use of plug-in estimators implemented as part of the JIDT toolbox \cite{lizier2014jidt}. For bias-correction, we used a Bayesian counting procedure implemented in the pyEntropy toolbox \cite{ince2009pyentropy}. For estimation of PID measures, we used the measure by Bertschinger et al. \cite{bertschinger2014unq} and an estimator by Makkeh et al. \cite{makkeh2017optimization,makkeh2018broja}, which is also part of the IDTxl toolbox. Analysis code is available from \cite{coderepo}.

\subsection*{Empirical data set}

We demonstrate the application of the proposed local information dynamics framework on spike train recordings from the retinogeniculate synapse of the cat. Spike trains were recorded from 17 retinal ganglion cells (RGCs) and monosynaptically coupled principal cells in the lateral geniculate nucleus (LGN) \cite{Rathbun2010}. We estimated $lAIS$ in the input to the synapse, i.e., the RGC spike train, and $lTE$ between the input and the output of the synapse, i.e., from the RGC to the LGN spike train  (Fig~\ref{fig:introduction}, \textbf{D}). We calculated LSTC to test whether information was preferentially transferred whenever the input signal was predictable or when it was non-predictable and we calculated the PID of the information transferred to validate our findings.

A detailed description of surgical procedures, task, and data recordings can be found in \cite{Rathbun2010}\footnote{Raw data are available from \url{https://github.com/scottiealexander/PairsDB.jl/raw/main/data/}.}. All surgical and experimental procedures were performed with the approval of the Animal Care and Use Committee at the University of California, Davis.

\subsubsection*{Surgery} For electrode placement at the RGC and LGN (Fig~\ref{fig:introduction}, \textbf{D}), adult cats of both sexes were initially anesthetized with ketamine (\SI[per-mode=symbol]{10}{\milli\gram\per\kilogram}, i.m.). For electrophysiological recordings, animals were placed in a stereotaxic apparatus and mechanically ventilated. Electrocardiogram (ECG), electroencephalogram (EEG), and expired \ce{CO2} were continuously monitored, while anesthesia was maintained with thiopental sodium (\SI[inter-unit-product = \ensuremath{{}\cdot{}}]{2}{\milli\gram\per\kilogram\per\hour}, i.v.). Thiopental administration was increased if physiological monitoring indicated a decrease in the level of anesthesia.

Once electrodes were positioned and minimum eye movement was ensured, the animal was paralyzed using vecuronium bromide (\SI[inter-unit-product = \ensuremath{{}\cdot{}}]{0.2}{\milli\gram\per\kilogram\per\hour}, i.v.). The pupils were dilated with \SI{1}{\percent} atropine sulfate and the nictitating membranes were retracted with \SI{10}{\percent} phenylephrine. Flurbiprofen sodium (\SI[per-mode=symbol]{1.5}{\milli\gram\per\hour}) was administered to ensure pupillary dilation. The eyes were fitted with contact lenses and focused on a monitor located \SI{1}{\meter} in front of the animal.

\subsubsection*{Visual task} Visual stimuli were created with a VSG 2/5 visual stimulus generator (Cambridge Research Systems) and presented on a gamma-calibrated Sony monitor with mean luminance of \SI[per-mode=symbol]{38}{\candela\per\square\metre}. Receptive fields were mapped using a binary white-noise stimulus that consisted of a \num{16} \texttimes \num{16} grid of squares \cite{Reid1997}. Each square flickered independently between black and white according to an \emph{m-sequence} \cite{Sutter1987,Reid1997}. The monitor ran at a frame rate of \SI{140}{\hertz}. Approximately \numrange{4}{16} squares of the stimulus overlapped the receptive field center of each neuron.

\subsubsection*{Electrophysiological recordings} Simultaneous single-unit recordings were conducted at the RGC and the contralateral layer A LGN cells. To maximize the chances that both cells were monosynaptically connected, a seven-channel multielectrode array (Thomas Recording) was placed in the LGN; through stimulation with a spot of light, the retinal area with the highest evoked response was identified. Cell responses were analyzed using an audio monitor.

Neural responses were amplified, filtered, and recorded with a Power 1401 data acquisition interface and the Spike 2 software package (Cambridge Electronic Design). The spikes of individual neurons were isolated using template matching, parametric clustering, and the presence of a refractory period in the auto-correlogram.

Recordings from \num{17} cell pairs entered further analysis. Recordings had an average length of \SI{788.4}{\second} ($\pm$ \SI{441.6}{\second} SD, see supporting information~S1).

To assert connectivity between recorded cells, the cross-correlogram between both recordings was visually inspected for abrupt, short-latency peaks using a bin-size of \SI{0.1}{\milli\second} (see \cite{Rathbun2010}, Fig 1). The occurrence of such a peak was seen as evidence for a monosynaptic connection between RGC and LGN cell \cite{Mastronarde1987,Usrey1998,Usrey1999}. For peaks a baseline mean was calculated from bins \SIrange{30}{50}{\milli\second} on either side of the peak bin. The peak bin and all neighboring bins with counts \textgreater\num{3}~SD were considered to contain retinal spikes triggering an LGN spike. The percentage of these spikes was termed the \textit{efficacy} of the RGC \cite{Levick1972,Usrey1998,Usrey1999}. Furthermore, an RGC's \textit{contribution} was defined as the percentage of LGN spikes that were triggered by a spike in the corresponding RGC. Contribution may be interpreted as the \enquote{strength of connection} between two cells in a pair \cite{Rathbun2010}. We called an RGC spike \textit{relayed} if it was followed by a LGN spike after its reconstructed information transfer delay, $u$ (see next section).

For further analyses recorded spike trains were binned into \SI{1}{\milli\second} segments.

\subsubsection*{Estimation of lAIS and lTE from empirical data}

We optimized nonuniform past-state embeddings for each cell pair recording using the greedy algorithm implemented in \cite{wollstadt2019idtxl}. For $lAIS$ estimation, we set the maximum lag, $j$, defining candidate variables for the embedding to \SI{30}{\milli\second}; for $lTE$ estimation, we set the maximum lag in the source, $k$, to \SI{40}{\milli\second} and the maximum lag in the target, $l$, to \SI{30}{\milli\second}. These lags assume that only spikes with an inter-spike interval (ISI) of \SI{30}{\milli\second} and less are relevant for triggering a LGN spike \cite{Mastronarde1987,Usrey1998,Levine2001,Rowe2001,Sincich2007,Weyand2007}, where especially ISIs \textless \SI{10}{\milli\second} are effective in driving LGN responses.

For optimizing the $lTE$ target past, we additionally account for a information-transfer delay, $u$, between RGC and LGN of up to \SI{10}{\milli\second}, which is in line with the cross-correlation observed between spiking in RGC and LGN \cite{Rathbun2010}. We reconstruct $u$ from the optimized embedding by identifying the lag of the past source variable that has the highest information contribution to the target's current state, quantified by the conditional mutual information $I(x_u: y_t|\mathbf{x}^S\setminus x_{t-u})$

\begin{equation}
	\hat{u} = \argmax_u I(y_{t}: x_{t-u} | \left\{\mathbf{y}^S, \mathbf{x}^S \setminus x_{t-u} \right\}) \, \forall x_{t-u} \in \mathbf{X}^s.
	\label{eq:delay_reconstruction}
\end{equation}

\subsubsection*{LSTCs calculation for empirical data}

To investigate whether the retinogeniculate synapse preferentially transferred predictable or unpredictable information, we correlated sample-wise estimates of $lAIS$ and $lTE$ by calculating the Pearson correlation coefficient between both measures. Note that we may also calculate measures that capture relationships of higher order, e.g., the mutual information. However, since our goal was to infer whether the sign of the correlation was positive or negative, we calculated the linear correlation. Tests for statistical significance were performed using a permutation test with \num{1000} permutations.

\section*{Results}

\subsection*{Optimization of estimation parameters}

For estimation of local information-theoretic measures, we first optimized past states for $lAIS$ and $lTE$ estimation individually for each cell pair. Over all cell pairs, the mean lag of variables identified for the $lAIS$ embedding was \SI{7.63}{\milli\second} (SD: \SI{1.82}{\milli\second}), and for $lTE$ embedding was \SI{2.75}{\milli\second} (SD: \SI{1.24}{\milli\second}) for the source and \SI{6.06}{\milli\second} (SD: \SI{1.53}{\milli\second}) for the target embedding. The reconstructed delay, $u$, between the RGC and LGN cell was on average \SI{2.81}{\milli\second} (SD: \SI{1.05}{\milli\second}), while individual delays matched maxima in the cross-correlogram between RGC and LGN recordings. (See supporting information~S1 for descriptive statistics of data entering the analysis, and supporting information~S2 for a list of all reconstructed parameters.)

\subsection*{LSTC}

\begin{figure}[!h]
	\includegraphics[width=\textwidth]{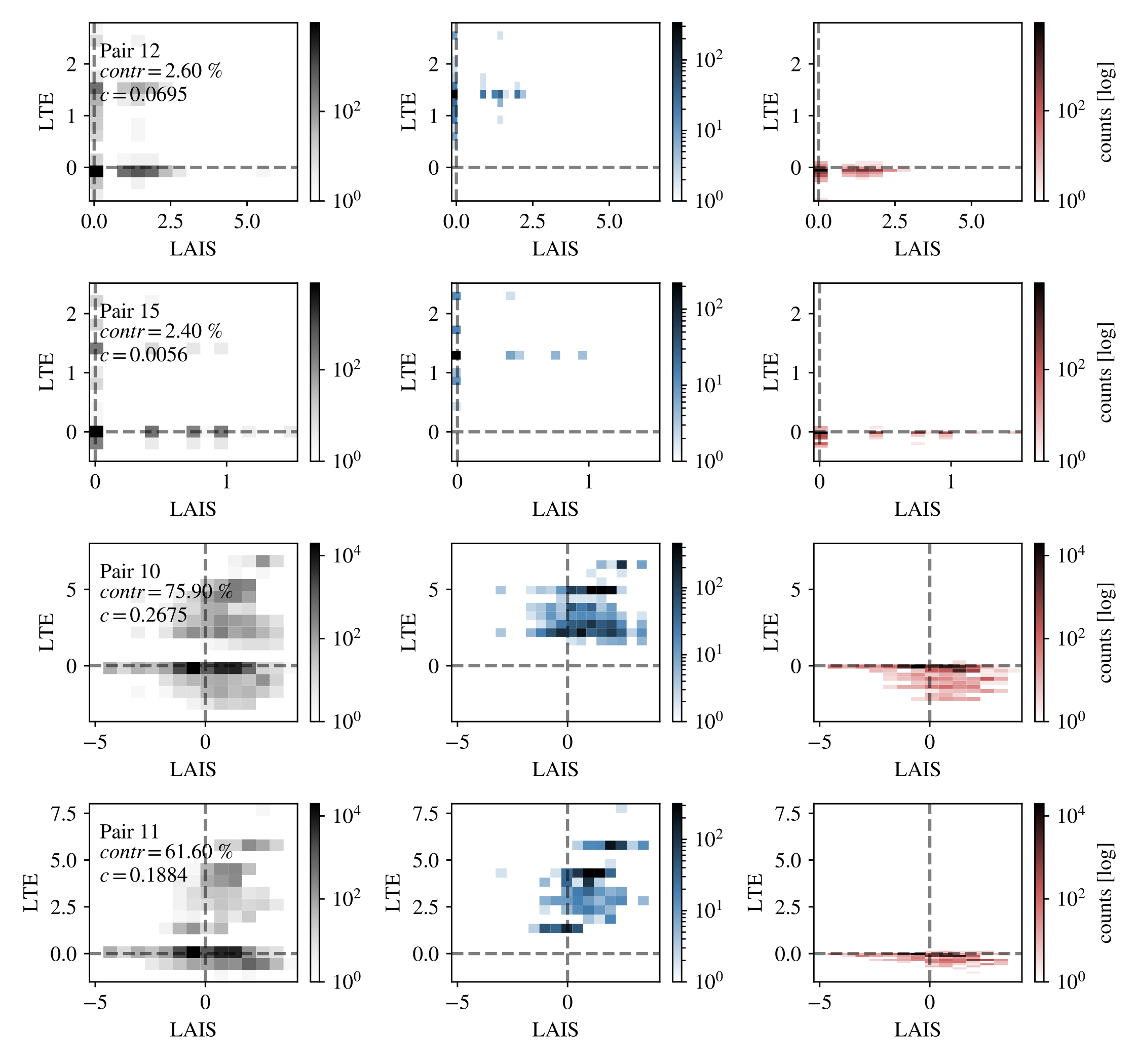} 
	\caption{{\bf Local storage-transfer correlations (LSTC) for exemplary cell pairs.}
		Histograms of LSTC for representative cell pairs with highest (pairs 10 and 11) and lowest (pairs 12 and 15) contribution, respectively. The first column shows histograms for all spikes, the second column for relayed spikes, and the third column for non-relayed spikes. Relayed spikes showed positive $lTE$ and generally positive $lAIS$, while non-relayed spikes led to zero or negative $lTE$ and lower $lAIS$.}
	\label{fig:lstc_corr_ex}
\end{figure}

Based on the optimized past states, we estimated $lAIS$ and $lTE$ for all cell pairs and found significant storage and transfer in all pairs except for pair \num{5}, which was excluded from all further analyses. For remaining cell pairs, we calculated the LSTC and found a significant, positive correlation coefficients for \num{14} of the remaining \num{16} pairs, indicating that local information transfer was higher at samples with higher local information storage (coefficients ranged from \numrange{0.0056}{0.2675}, see also supporting information~S3). With respect to predictive coding strategies, the positive LSTC indicates a higher transfer of information whenever an input sample was more predictable from its past, and less transfer when it was unpredictable. Fig~\ref{fig:lstc_corr_ex} shows two-dimensional histograms of $lAIS$ and $lTE$ values for RGC spikes for four representative cell pairs.

\begin{figure}[!h]
    \centering
	\includegraphics{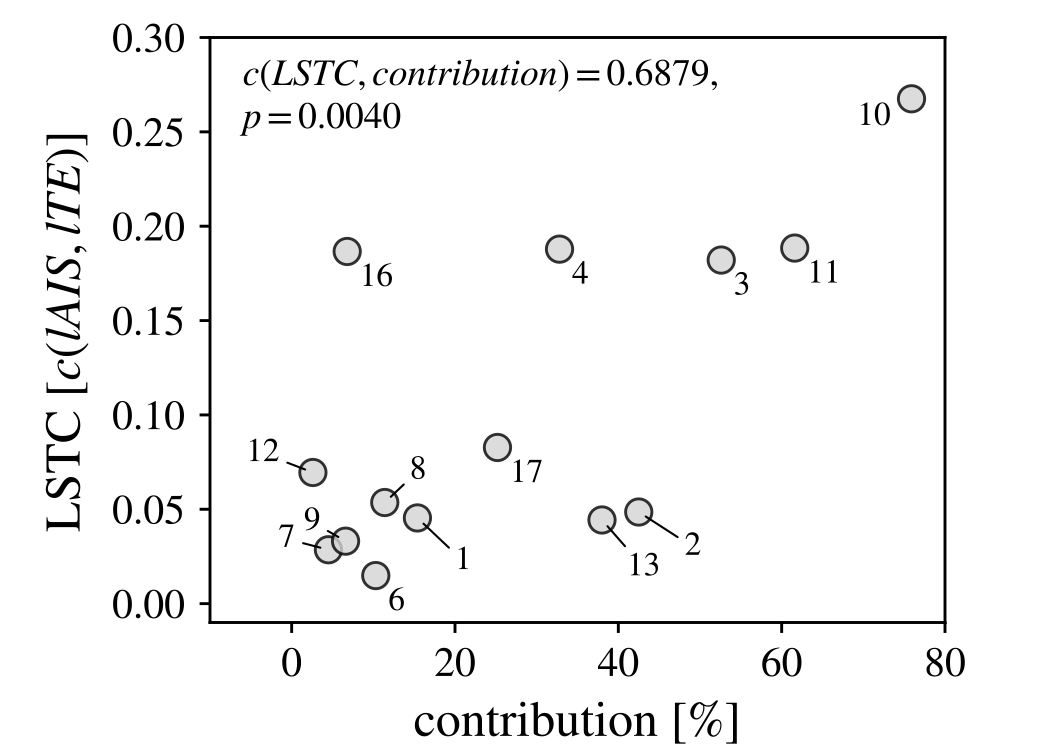}  
	\caption{{\bf Correlation between contribution and local storage-transfer correlations (LSTC) for all spike pairs.}}
	\label{fig:lstc_contribution}
\end{figure}

We further found that correlations were stronger in cell pairs with a high RGC \emph{contribution} (Fig~\ref{fig:lstc_contribution}, $c(LSTC, contribution)=0.6879$, $p = 0.0030^{**}$). In a cell pair, the RGC's contribution is defined as the percentage of spikes in the LGN cell that were triggered by a previous spike in the RGC and may be interpreted as the pair's \enquote{strength of connection} \cite{Rathbun2010}. Hence, the effect of predictable information being relayed across the retinogeniculate synapse was more pronounced in synapses that were more strongly connected. Estimates of the contribution of each cell pair were taken from \cite{Rathbun2010}.

\subsection*{Information dynamics of relayed and non-relayed RGC spikes}

For all observed pairs, typically only a fraction of RGC spikes was \emph{relayed} to the LGN, i.e., followed by an LGN spike with the reconstructed information transfer delay, $u$. Remaining RGC spikes were considered \emph{non-relayed}. We investigated whether such relayed RGC spikes differed in their local information dynamics from non-relayed spikes.

\begin{figure}[!h]
	\includegraphics[width=\textwidth]{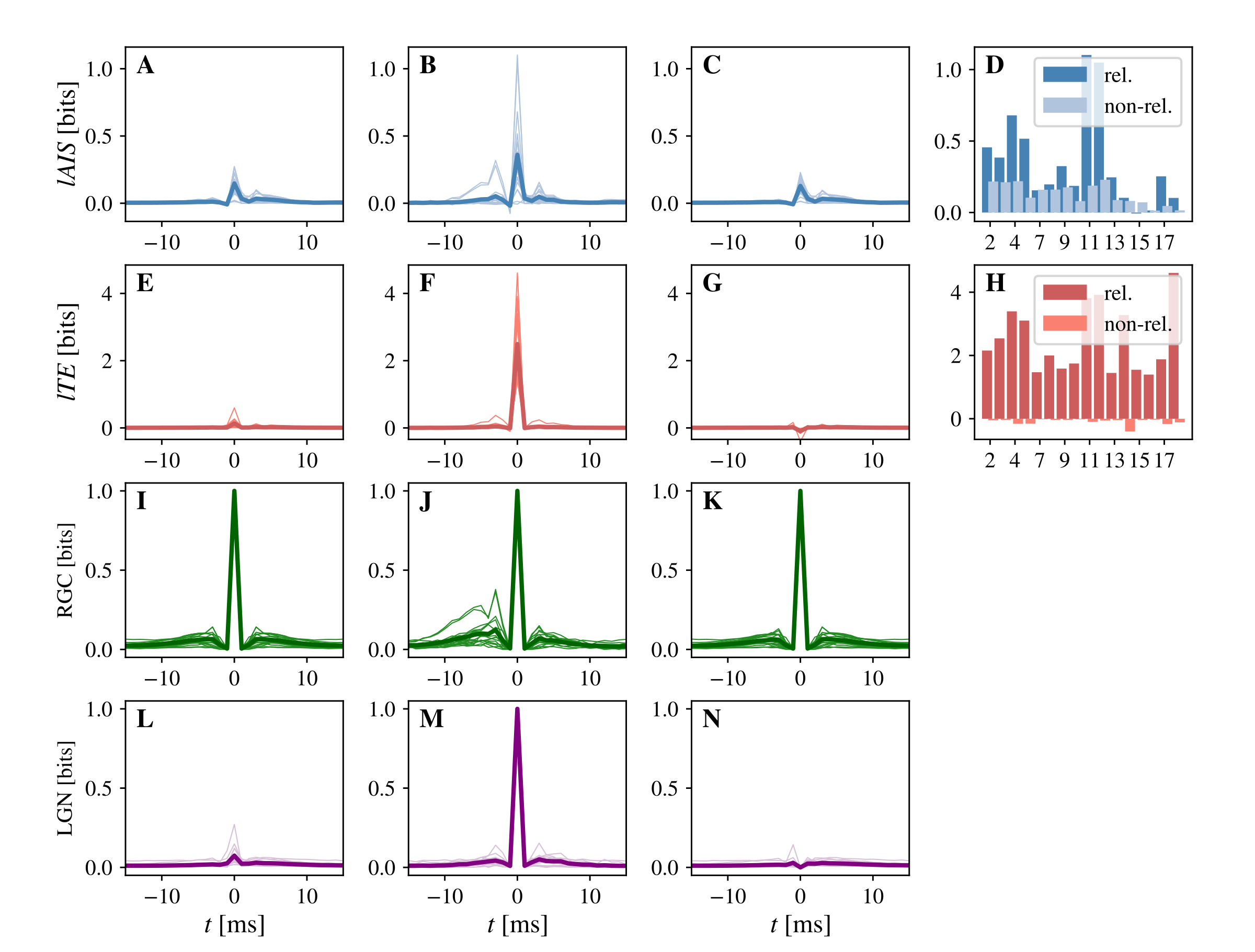} 
	\caption{{\bf Information dynamics of relayed versus non-relayed RGC spikes.}
		An RGC-spike was considered relayed to the LGN if it was followed by an LGN spike with the delay reconstructed as part of $lTE$ estimation.
		Spike-triggered average (STA) for $lAIS$ values for (\textbf{A}) all, (\textbf{B}) relayed and (\textbf{C}) non-relayed RGC-spikes; \textbf{D} $lAIS$ values for each cell pair at relayed (dark blue) and non-relayed (light blue) RGC spikes ($p < 0.001^{***}$ for a permutation test with \num{1000} permutations).
		STA for $lTE$ values for (\textbf{E}) all, (\textbf{F}) relayed and (\textbf{G}) non-relayed RGC-spikes; \textbf{H} $lTE$ values for each cell pair at relayed (dark red) and non-relayed (light red) RGC spikes ($p < 0.001^{***}$ for a permutation test with \num{1000} permutations).
		STA of LGN spike train for (\textbf{I}) all, (\textbf{J}) relayed, and non-relayed (\textbf{F}) non-relayed RGC-spikes;
		STA of RGC spike train for (\textbf{L}) all, (\textbf{M}) relayed, and non-relayed (\textbf{N}) non-relayed RGC-spikes.
    }
	\label{fig:sta}
\end{figure}

On average, relayed RGC spikes were accompanied by higher $lAIS$ and $lTE$ compared to non-relayed spikes (Fig~\ref{fig:lstc_corr_ex} and Fig~\ref{fig:sta}). Results indicate that, first, relayed spikes were in general more predictable from the RGC's cells immediate past spiking behavior. Second, relayed spikes were accompanied by higher local information transfer, while non-relayed spikes were accompanied by negative local information transfer. Negative $lTE$ here means that for some cell pairs, in the absence of an LGN spike, the RGC's state (spike) was misinformative about the next state of the LGN (no spike). In other words, observing a prior RGC spike lowered the probability of observing no spike in the LGN.

Relayed spikes were characterized by both higher $lAIS$ and $lTE$ values. We were able to classify whether a spike was relayed from its $lAIS$ value above chance, using a k-nearest neighbor classifier with $k=5$ (classification accuracy was also higher than the baseline model, see supporting information~S2). However, note that $lAIS$ may be seen as a different representation of spiking statistics of the RGC, i.e., the number of spikes and ISI in a given time window (as can be seen, for example, when considering spike-triggered averages of $lAIS$ and RGC spike counts in Fig \ref{fig:sta}). As a result, whether a spike was relayed could be equally well predicted from the spike count of all spikes up to \SI{30}{\milli\second} prior to an RGC spike, or the RGC spike's prior ISI (see supporting information~S2). We therefore want to highlight that the estimation of $lAIS$ provides no additional, mechanistic explanation on \textit{when} a spike is relayed at the retinogeniculate synapse---it rather provides a computational interpretation of the mechanisms already known (see Discussion).

\subsection*{Information dynamics of inter-spike intervals}

\begin{figure}[!h]
	\includegraphics[width=\textwidth]{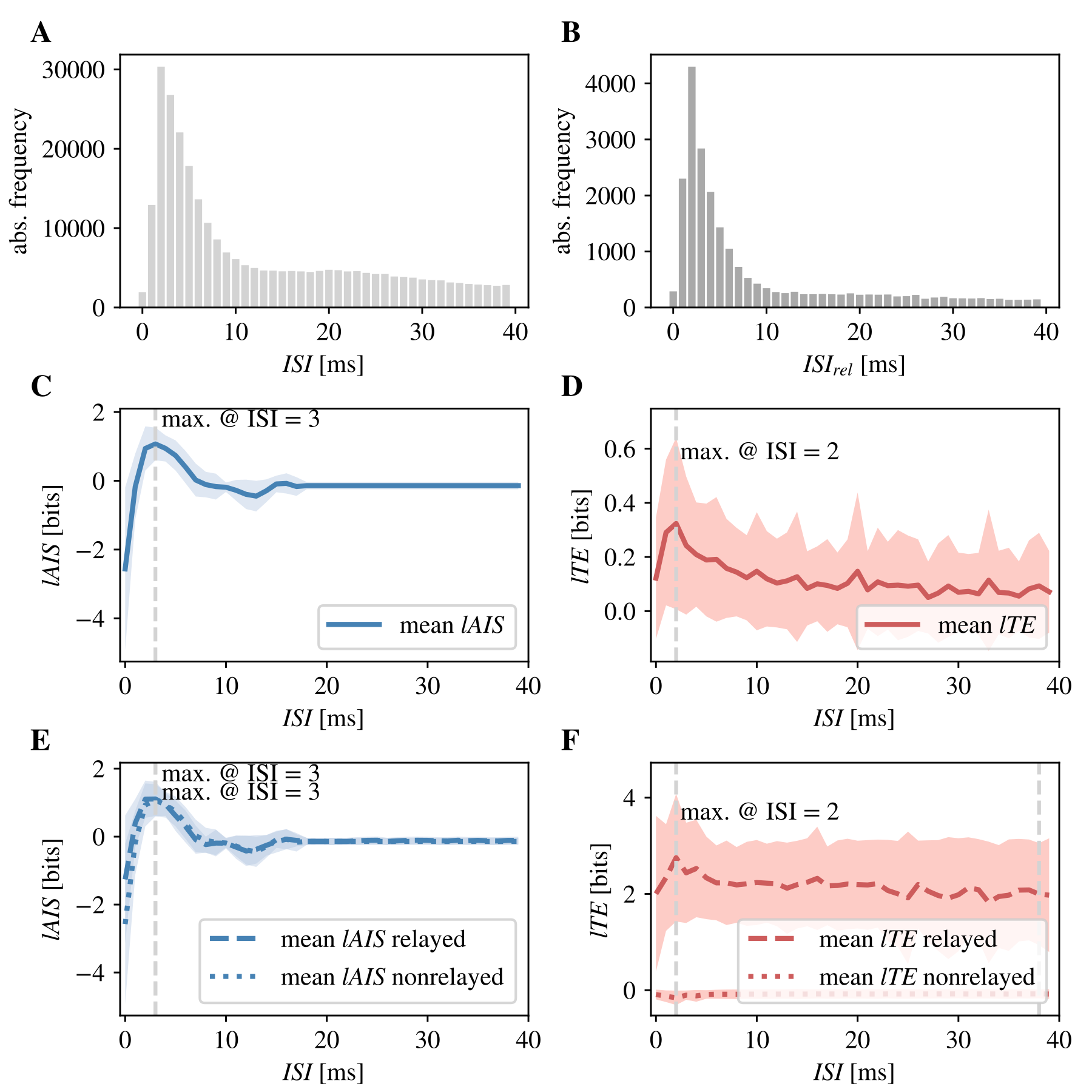} 
	\caption{{\bf Information dynamics of inter-spike intervals (ISI).}
		\textbf{A} Distribution of ISI pooled over all cell pairs (maximum at \SI{3}{\milli\second}, median of \SI{26.65}{\milli\second}, and standard deviation of \SI{56.80}{\milli\second}).
		\textbf{B} $lAIS$ at RGC spike as a function of the preceding ISI (maximum at ISI= \SI{3}{\milli\second}, dashed vertical line);
		\textbf{C} $lTE$ at RGC spike by preceding ISI (maximum at ISI= \SI{2}{\milli\second}, grey vertical line);
		\textbf{D} $lAIS$ at relayed (dotted line) and non-relayed (dashed line) RGC spikes as functions of the preceding ISI (maxima at ISI= \SI{2}{\milli\second} for relayed spikes, dotted vertical line, and at ISI=\SI{3}{\milli\second} for non-relayed spikes, dashed vertical line);
		\textbf{E} $lTE$ at relayed (dotted line) and non-relayed (dashed line) RGC spikes as functions of the preceding ISI (maxima at ISI= \SI{2}{\milli\second} for relayed spikes, dotted vertical line, and at ISI=\SI{28}{\milli\second} for non-relayed spikes, dashed vertical line).
		}
	\label{fig:isi}
\end{figure}

Next, we investigated the local information dynamics of RGC and LGN spikes as a function of the preceding ISI, as ISIs have been reported to have an effect on whether an RGC spike drives a response in the corresponding LGN cell \cite{Mastronarde1987,Usrey1998,Levine2001,Rowe2001,Sincich2007,Weyand2007}. We calculated ISIs by subtracting the spiking times of all consecutive spikes in the RGC spike train (Fig~\ref{fig:isi}\textbf{A}).

Average $lAIS$ was positive for RGC spikes with a preceding ISI of \SIrange{2}{7}{\milli\second}, with a maximum at \SI{3}{\milli\second}. The $lAIS$ was negative for all other investigated ISIs (Fig~\ref{fig:isi}\textbf{B}). Hence, the most frequent ISIs lead to higher predictability of the spike. When differentiating between relayed and non-relayed RGC spikes, $lAIS$ was positive for ISI of \SIrange{1}{6}{\milli\second} for relayed spikes while the range of positive $lAIS$ values for non-relayed spikes was \SIrange{2}{7}{\milli\second}. Overall, relayed and non-relayed spikes did not differ in $lAIS$ as a function of ISI (Fig~\ref{fig:isi}\textbf{D}). $lTE$ was positive over the whole range of investigated ISIs (Fig~\ref{fig:isi}\textbf{C}). However, when differentiating between relayed and non-relayed spikes, $lTE$ was negative on average for all ISIs for non-relayed spikes with a minimum at \SI{2}{\milli\second} (Fig~\ref{fig:isi}\textbf{E}).

We further investigated RGC spike \emph{tuples}, because whether RGC spikes are relayed is mostly influenced by the most recent previous spike while events further in the past have only minor influence \cite{Usrey1998}. Tuples are defined as two spikes with an ISI below a given threshold and a \enquote{silence time} preceding the first spike to ensure a comparable level of prior activity \cite{Usrey1998}. We here used a silence time and maximum ISI of \SI{20}{\milli\second}, which covers the maximum history length used in the estimation of $lAIS$ and $lTE$ (supporting information~S1), such that spikes with a prior ISI \textgreater \SI{20}{\milli\second} did not influence $lAIS$ and $lTE$ estimates.

\begin{figure}[!h]
	\includegraphics[width=\textwidth]{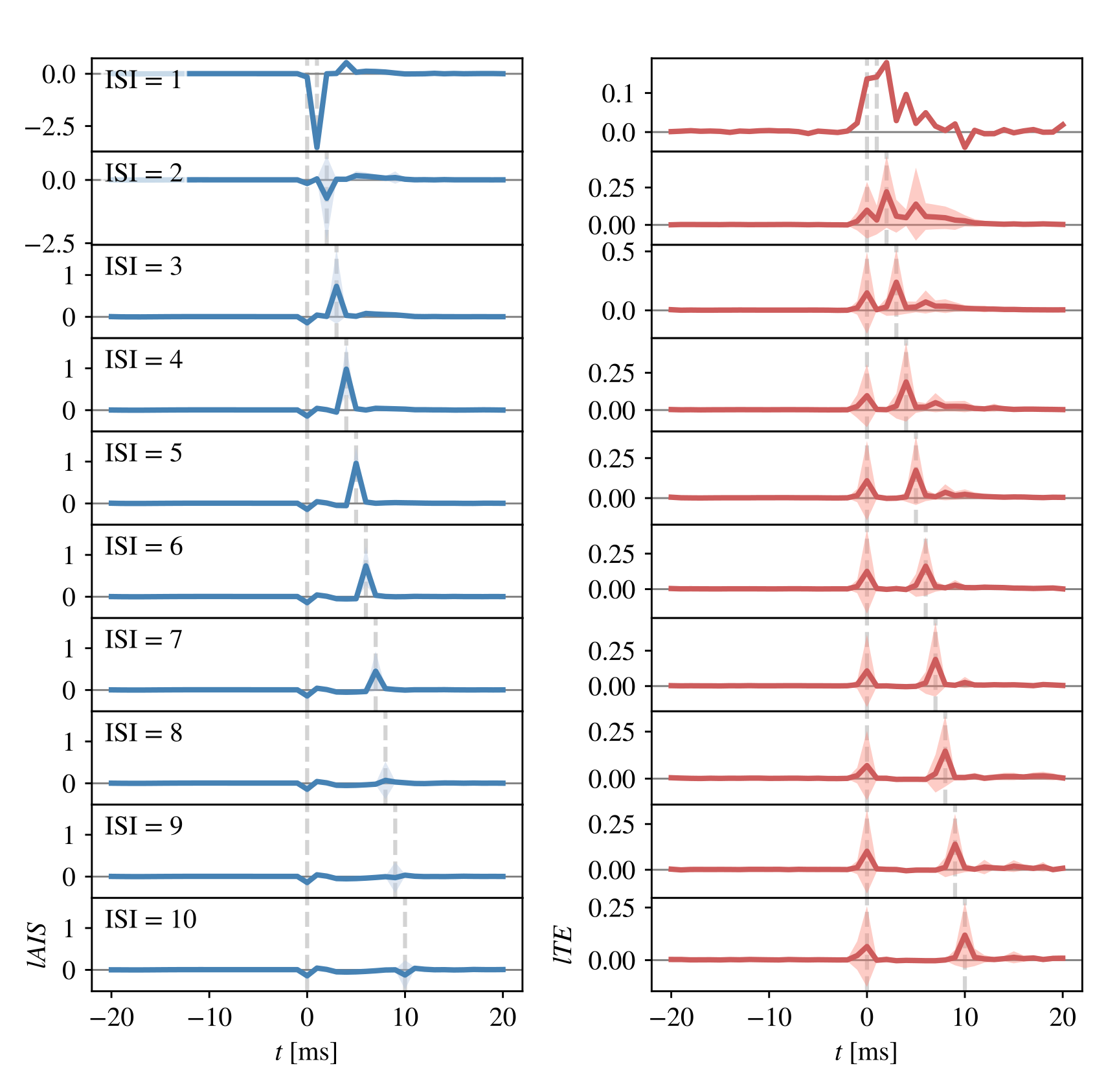}  
    \caption{{\bf Spike-triggered averages (STAs) for spike tuples.}
	STAs for spike tuples with a silence time of \SI{20}{\milli\second} and inter-spike interval (ISI) up to \SI{20}{\milli\second} (aligned on first spike in a tuple). Left column shows $lAIS$ values averaged over cell pairs for ISI of \SIrange{1}{10}{\milli\second}, right column shows averaged $lTE$ values (shaded areas indicate $\pm 1 SD$). Note that $lTE$ values are shifted by the individual delay between RGC and LGN cell for each pair. Hence, a spike at index $t=0$ indicates a transferred spike with a delay corresponding to the reconstructed information transfer delay.}
	\label{fig:tuples}
\end{figure}

We computed spike triggered averages (STA) of $lAIS$ and $lTE$ values for spike tuples (two consecutive spikes with an with an ISI \textless \SI{20}{\milli\second}, Fig~\ref{fig:tuples}). On average, the first spike in a tuple was associated with negative $lAIS$, indicating that the spike's immediate past, i.e., the silence time, was misinformative about the spike. For an ISI of \SIrange{3}{8}{\milli\second}, the second spike was associated with increased $lAIS$, relative to the average $lAIS$ in the silence time, indicating high predictability from the immediate past.
On average, $lTE$ values were slightly increased for the first and second spike in a tuple, with higher values for the second spike. In sum, the predictability of an RGC spike strongly dependent on prior spiking activity, with higher predictability if the ISI was between \SI{3}{\milli\second} and \SI{8}{\milli\second}.

\subsection*{State-dependent and state-independent information transfer}

\begin{figure}[!h]
	\includegraphics[width=\textwidth]{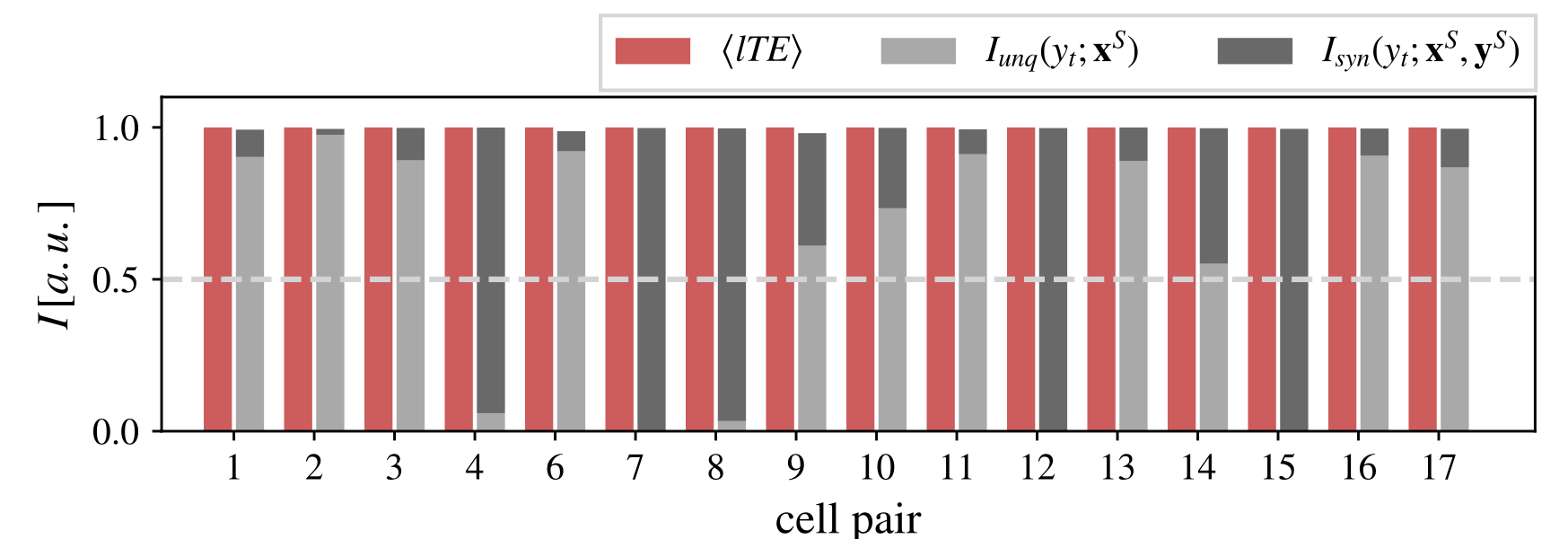}  
	\caption{{\bf State-dependent and -independent information transfer from RGC to LGN cell.}
		State-dependent and -independent information transfer from RGC to LGN cell, measured by the synergistic information, $I_{syn}(Y_t;\mathbf{X}^S, \mathbf{Y}^S)$ (dark gray), and unique information, $I_{unq}(Y_t;\mathbf{X}^S)$ (light gray). In 11 out of 16 pairs, more than half of the transferred information was predominantly state-independent.}
	\label{fig:pid_contribution}
\end{figure}

Last, we estimated unique, $I_{unq}(Y_t:\mathbf{X}^S)$, and synergistic information, $I_{syn}(Y_t;\mathbf{X}^S, \mathbf{Y}^S)$, the RGC's past provided about the spiking behavior of the LGN principal cell, using the measure by Bertschinger et al. \cite{bertschinger2014unq,makkeh2018broja} (Fig~\ref{fig:pid_contribution}). In \num{11} out of \num{16} cell pairs with significant information transfer, the unique information provided by the RGC's past state, $\mathbf{X}^S$, dominated the information transfer from RGC to LGN. Hence, information transfer was governed by information transfer independent of the state of the LGN's past state. Again, this supports the notion of information transferred mainly in a bottom-up fashion, i.e., transfer independent of the target cell's state.

\section*{Discussion}

We introduced an information-theoretic framework for testing predictive coding strategies in neural data. The framework expresses predictive coding concepts, namely predictability, predictions, and prediction errors, in terms of information-theoretic quantities, which can be immediately estimated from data. Hence, the framework does not rely on markers of predictive coding that have to be defined a priori, but on properties of the data itself. As a result, the framework is able to investigate neural processing strategies in any data set, independently of the experimental task under which the data were collected.

We applied the framework to spike recordings from the retinogeniculate synapse of the cat and identified the preferred coding strategy of the synapse, namely the transfer of predictable over unpredictable input. In particular, we showed that the RGC-LGN synapse preferentially transferred information whenever the input from retinal cells to LGN cells was highly predictable from its immediate past. This conclusion is supported by the finding that the transferred information was predominantly bottom-up input to the synapse, and was independent of the state of the LGN principal cell.

\subsection*{Local information dynamics as a semantics-free approach to investigating neural computation}
\subsubsection*{Avoiding circular arguments in the investigation of predictive coding theory} The aim of this study was to develop an information-theoretic framework that allows to test predictive coding theories in arbitrary recordings of neural data, that is, recordings that were performed in setups not specifically tailored to investigate predictive coding. One key motivation for such a framework was to avoid the use of circular arguments where the researcher's assumptions on what inputs the brain should predict are used to design stimuli and paradigms, that are then used in neurophysiological experiments to test whether and how the brain predicts these inputs. As a result, the experimenter's interpretation of neural activity becomes dependent on the a priori defined theory motivating the experimental setup and the expected manifestation of a prediction error in the data (see \cite{kogo2015predictive,walsh2020evaluating,DeWit2010,litwin2020unification} for other descriptions of this problem). This motivation is based not least in the difficulties we experienced ourselves when designing and performing predictive coding experiments, and interpreting the data (e.g. in \cite{brodski2015faces}). However, we do not want to suggest that all experiments necessarily suffer from these difficulties. Rather, we fully acknowledge that in some cases the necessary knowledge to carefully design non-circular predictive coding experiments will be available. However, we are concerned that this is not the case in general.

The problem of \enquote{interpreting} neural activity in terms of predictions or prediction errors becomes even more severe if arbitrary processing elements in the cortex are investigated in isolation, e.g., single cells, whose computations, as well as input and output are far removed from any human-understandable function (i.e. we have to treat it as  \enquote{intrinsic computation} in the sense of \cite{Mitchell1993edgeofchaos}).
Here, an approach is required that analyzes signals not from an \enquote{experimenter-as-receiver}, but a \enquote{cortex-as-receiver} or \enquote{neuron-as-receiver} point of view \cite{DeWit2016}. The former view assumes that neural signals at arbitrary processing stages carry human-understandable information---which may express a misleading view on information processing in the brain in general---, while the latter view considers the question of how other processing units in the cortex view available information.

Here, local information dynamics provides a tool that allows for the direct investigation of computations performed by arbitrary processing units. By expressing the building blocks of predictive coding theories, namely, prediction, information transfer, and prediction errors, in terms of information-theoretic quantities, these concepts become measurable properties of the data such that we were able to formulate testable hypotheses on how these quantities should relate given two opposing predictive coding strategies. Our approach thereby did not rely on an interpretation of the data recorded in terms of the experimental task performed (compare, for example, previous studies investigating the facilitation of predicted input over the propagation of prediction errors \cite{Roelfsema1998,Sillito1994,Luck1997,Desimone1995}). 
Rather, using information-theoretic measures allowed us to \enquote{abstract away} from the experimental setup such that predictive coding could be investigated independently of the task. The presented approach not only allows for a straightforward testing of competing information-processing strategies in arbitrary neural systems, but also opens up the possibility of testing predictive coding theories on data from a myriad of neurophysiological experiments not initially designed with predictive coding in mind.

We acknowledge that the proof-of-principle analysis presented here is an extreme case of application: The random stimuli lack all predictability, except for the short life-time of a video-frame. Thus, any predictability in the spike time series is internally generated in the retina. As a consequence, one may ask whether our analysis---although technically sound---is truly relevant, and whether the example analyzed here supports a more general applicability. For two reasons we think this is indeed the case---first, $lAIS$ must certainly further rise in all processing stages close to a predictable stimulus---compared to the unpredictable one used here, thus increasing the signal-to-noise ratio relevant for its estimation; second for a neuron receiving a spike train, it will essentially be difficult to distinguish internal from external sources of predictability in that spike train---at least without resorting to information from other (neural) channels. This latter problem of exploiting information from additional channels is indeed important in predictive coding and discussed next. How such side-channel information can be integrated in our analysis framework is discussed further below.

\subsubsection*{Quantifying the predictability of neural signals as a proxy for predictions}
When investigating predictive coding strategies at the retinogeniculate synapse, we quantified the \textit{self-predictability} of the input signal to the synapse in order to quantify what portion of the signal was predictable from its own immediate past. We used this predictability as a proxy for measuring and quantifying an actual prediction of the synapse's input from all available inputs (e.g., through feedback connections from the cortex to the LGN principal cell etc.). Using the self-predictability as a proxy for quantifying predictions introduces three approximations:

The first approximation is that a mutual information is used in place of an actual predictive model for the RGC inputs embodied in the organism. This mutual information is an upper bound on the mutual information between an actual model prediction based on the RGC inputs and the realized future samples of a time series. The second approximation introduced here relates to the practical estimation of the local active information storage from spike trains. With limited data, this can lead to biased results, either by overestimation for the case of severely limited data availability or by underestimation if the analyzed history length is artificially shortened to curb 'curse of dimensionality' problems in the estimation (see also \cite{rudelt2021embedding} for a discussion of these issues). Given the large amount of data available here we do not consider this limitation to apply. Last, a third approximation is introduced by looking only at the information available for prediction in the input spike train of the RGC cells. It is obviously conceivable that neurons from other brain regions that have access to information arising from multiple retinal inputs across a wider range of visual field locations could potentially improve a prediction above what is predictable from the RGC input itself. This is indeed a valid concern. Below, we present a way of incorporating such information into the estimation of predictable information. In the present experiment such additional influences, e.g. from cortical feedback, will possibly be negligible do to the anesthetized state of the animal, and the random nature of the stimulus. In other scenarios, information from cortical channels, or other side channels, may have to be incorporated using Eq~\ref{eq:V1sidechannel} below.

This first assumption holds under a second assumption, namely, that we choose the past state of the input used in the estimation of AIS such that we indeed capture all relevant statistical regularities. In theory, the AIS is defined as the mutual information between a processes' present state and its \textit{semi-infinite} past \cite{lizier2012lais}. Hence, in practice one has to find a suitable, finite embedding that covers only the \textit{relevant} past \cite{ragwitz2002,lizier2012lais,faes2011nonuniform}. Such an embedding is found through the approach used here, where we optimized a non-uniform embedding that covered a time horizon of \SI{30}{\milli\second}, which was identified in previous work as the time horizon over which spikes affect future spiking behavior of the RGC \cite{Mastronarde1987,Usrey1998,Levine2001,Rowe2001,Sincich2007,Weyand2007}.

\subsubsection*{Quantifying prediction errors in neural signals} We can use predictability not only as a proxy for the prediction of an input signal, but also as a proxy for \emph{inevitable} prediction errors: If predictability of the input signal is low, and thereby its $lAIS$,---according to our first assumption above---any reasonable model predicting the synapse's next state from this input must generate a prediction error---simply by virtue of reflecting the underlying probabilities.

As a second approach to quantifying prediction errors, we proposed to calculate the synergistic portion of the information transfer between the synapse's in- and output, using the recently proposed PID framework. In particular, we propose that high synergistic information between the input-cell's past state and the target-cell's past state about the target-cell's next state reflects transmission of a prediction error. This is because both, the past state of the input cell, providing the sensory evidence, and the past state of the target cell, providing the prediction, have to contribute to the computation of the target cell's next state, if this state reflects a prediction error. In contrast, transfer of primarily bottom-up, sensory evidence is indicated by high unique information in the input cell's past state about the target cell's next state, while the synergy between both past states is low (this is discussed in more detail in section \textit{Predictive coding at the retinogeniculate synapse} below).

\subsubsection*{Quantifying information transfer between neural signals}
Next to predictability and prediction errors, we estimated information transfer across the synapse using TE. TE quantifies how much information is transferred from an input to an output variable. Thus, TE serves as a natural measure of information transfer serving predictive coding, i.e., the transfer of \textit{novel} information from input to output of the receiving cell. As discussed in the previous section, the information transfer across the synapse, quantified by the TE, comprises both information uniquely provided by the synapse's input, but also information transferred due to synergistic effects between past state of the input and past state of the target cell.

Next, we will review how our information-theoretic results relate to possible biophysical implementations of the computations performed.

\subsection*{Linking information-theoretic results to biophysiology}

We found that information transfer was highest for highly predictable RGC spikes and that these input spikes were typically preceded by another input spike with a short advance. Our findings are in line with previous studies showing that RGC spikes with a preceding RGC spike were more effective in driving an LGN response than single spikes \cite{Usrey1998,Rowe2001,Rathbun2007,Weyand2007,Rathbun2010}, and that this efficacy was even higher for ISI \textless \SI{10}{\milli\second}  \cite{Cleland1971,Usrey1998,Levine2001a,Rowe2001,Rathbun2007,Sincich2007,Weyand2007}.

It has been hypothesized that double spikes are important to enable temporal summation at the post-synaptic membrane: Carandini and colleagues \cite{Carandini2007} presented a model of the retinogeniculate synapse in which information transfer was governed by temporal summation of pre-synaptic excitatory postsynaptic potentials (EPSP). Here, EPSPs remained approximately constant or even increased for smaller ISI. Hence, the dominant biophysical mechanism enabling information transfer at the synapse seemed to be post-synaptical summation rather than a change in pre-synaptic conditions due to enhanced spike-rates. This fits with the LGN's limited ability to integrate spikes over large time windows, where the typical time constant for X- and Y-cells in the LGN is measured to be \SIrange{15}{22}{\milli\second} \cite{Crunelli1987}. The cells' true ability to perform temporal summation may be even lower because the time constant may not be a suitable measure of the ability for temporal integration under real-world conditions \cite{Koch1996}. Temporal summation as a mechanism is further supported by the fact that LGN principal cells receive input from just a small number of RGCs of which one is typically the main driver \cite{Reid2004,Sincich2007}, furthermore, also single RGC cells are able to drive the target LGN principal cell \cite{Cleland1971,Usrey1998,Levine2001,Rowe2001,Weyand2007}, such that population coding is an unlikely mechanism for the information transfer at the retinogeniculate synapse. Also---as was noted by Rowe and colleagues---contribution rises strongly under \enquote{structured} visual stimulation \cite{Rowe2001}.

In sum, temporal summation over incoming spiking activity on short time-scales is a likely mechanism for driving information transfer from RGC to LGN principal cells. Our findings are compatible with this mechanism, as information transfer was highest for the second RGC spike in tuples with short ISI (see Fig. \ref{fig:tuples}). This spike also had high predictability, explaining the observed correlations based on the above biophysical mechanism. Furthermore, when applying partial information decomposition we found predominantly unique information transfer from RGC to LGN, which is in line with the fact that almost all LGN spikes are triggered by an RGC spike \cite{Cleland1971,Rathbun2010}. In sum, our novel framework yields results that have a mechanistic explanation here in the observed properties of biological neurons (and it would be a reason for concern if this were not the case), but it adds another explanatory layer to the mere biophysiological description. It does so by measuring exactly what the \emph{information processing consequences} of the established biophysical principles are. This computational description in information-theoretic terms allows us to bridge the explanatory gap between biophysics and predictive coding theories. Also, our information-theoretic analysis shows that basic predictive coding for reliable information (i.e. for predictable inputs) may in some cases be realizable by cellular biophysical principles alone, if input spiking statistics allow for exploiting these principles. It is an open question, however, whether cellular biophysics alone could also be exploited for coding for prediction errors.

\subsection*{Predictive coding at the retinogeniculate synapse}

We applied the proposed local information dynamics framework to investigate which of two alternative predictive coding strategies were used at the retinogeniculate synapse. In particular, we tested, whether the synapse coded for unpredictable or surprising input versus predictable input. Both coding strategies have been formulated as part of wider theories of predictive coding. The first strategy is proposed by, for example, \cite{Rao1999} or \cite{Mumford1992,Hohwy2008}, and states that bottom-up signals in the \emph{cortical} hierarchy generally reflect the propagation of prediction errors. This family of predictive coding theories also proposes that the top-down signals represent predictions made at a higher cortical area about the next lower area in order to \enquote{explain away} sensory input at the lower area \cite{Rao1999,Mumford1992,Hohwy2008}. Here, the bottom-up error-signal represents the part of the top-down prediction not explained away and thus signals the mismatch between prediction and input \cite{Friston2010gut}. The second coding strategy is proposed as part of a further family of predictive coding-like theories, which oppose the propagation of prediction errors and instead state that the bottom-up signal in the cortical hierarchy represents predictable input. Examples of this family are, amongst others, ART \cite{Grossberg1980,Grossberg2013} and the biased competition model \cite{Desimone1995,Desimone1998}. Both theories assume that sensory input matching top-down information is amplified and propagated up the cortical hierarchy.

Currently, it is an ongoing debate which of the two proposed strategies neural systems use. Both strategies have been shown to be equivalent on a \emph{functional level} (they use the principle of predictive coding to realize perception and action in the cortex), while they differ  on an \emph{algorithmic level} \cite{Marr1982}, and as a consequence in their implementation. Spratling and colleagues showed that both strategies are equivalent in their ability to realize predictive-coding-like information processing in artificial neural networks \cite{Spratling2008}. This was supported by Kveraga et al. \cite{Kveraga2007}, who suggest that different realizations of PCT, such as biased competition model, ART, and error-coding theories, could be easily accommodated by a computational model of top-down and bottom-up information processing presented in \cite{Siegel2000} (see also \cite{Summerfield2009} for a further comparison of theories on top-down activity). Here, our framework provides an alternative approach to testing neural coding strategies against each other.

\subsubsection*{Evidence for coding for predictable input found at the retinogeniculate synapse}

Applying the proposed framework, we found that the retinogeniculate synapse preferentially coded for predictable input. In \num{15} of \num{17} investigated cells, local predictability of the input correlated positively with local information transfer between input and output indicating the preferential transfer of predictable input. Also, we found that RGC spikes were more efficient in driving an LGN response if they were highly predictable from their immediate past. Lastly, we used PID \cite{Williams2010,bertschinger2014unq} to decompose information transfer from the RGC to the LGN principal cell into information uniquely provided by the RGC about the next state of the LGN and into synergistic information provided jointly by the RGC's and the LGN's past. We found that unique information in the RGC was transferred in the majority of investigated cell pairs, opposed to the transfer of synergistic information, which indicates that primarily sensory evidence was transferred across the synapse.

The last finding on information transfer being dominated by unique information being relayed from the RGC supports a coding strategy for predictive information as follows: if the information transfer across the synapse served the propagation of prediction errors, we would expect high synergistic information transfer across the synapse. This is due to the fact that to calculate the occurrence of a prediction error, both, the past state of the target cell---the prediction---and the past state of the input cell---the sensory evidence---had to be known. The error would then be computed from comparing these two inputs, leading to a response if there was a mismatch between the two states. Technically, for single spiking events, perfectly determining the occurrence of a prediction error is equivalent to a binary XOR operation, which leads to purely synergistic information between the two inputs and the output. That is because knowing only one input does not provide any information what the output should be (see also the next section). While it is also well known that single neurons can only approximate a binary XOR, this would still lead to considerable synergistic information. Conversely, if the information transfer across the synapse served the propagation of predictable information, we would expect low synergistic information and some unique information in the input about the output (in a process similar to a binary AND operation). The latter scenario is in line with our empirical findings, indicating that the input to the synapse provided unique information about the next state of the target cell (observing a spike in one source increases the probability of observing a spike in the output). Hence, PID results support the coding for predictable input rather than coding for prediction errors.

Lastly, we want to emphasize that the framework presented here does not provide us with new information about the processing at the synapse, i.e., information that is not already contained in the spiking statistics of the cell. Rather, the framework is an approach that---while being task-agnostic---is able to cast computations performed by the biophysical dynamics into a quantitative and human-interpretable form. Indeed, our finding of a preferential transfer of predictable input sheds an interesting light on the findings in \cite{Rathbun2010}. Predictable input to the LGN cells (spike tuples) is produced when an RGC cell is stimulated by its preferred input. Thus, the signals relayed by LGN cells are strongly representational in nature, rather than differential. In sum, the biophysical mechanism and its function in terms of enhancing representations was known, but our analysis adds a quantitative computational interpretation within a framework that can be applied to radically different circuits for comparison.

\subsubsection*{Potential influences of anesthesia on cortical feedback and predictive coding}\label{CorticalSidechannel}

In our analysis, we used recordings from animals under anesthesia, which may affect our results due to the well-known change in information transfer under anesthesia, predominantly in top-down direction \cite{Imas2005,Ku2011STE,Lee2013,Jordan2013,Untergehrer2014,Wollstadt2017}. Hence, under anesthesia top-down information transfer from V1 to the LGN is very likely reduced and it is conceivable that LGN function---and thus the algorithm embodied by the synapse---differs between the anesthetized and awake state. V1 activity affects LGN function via direct and indirect connections (e.g. \cite{Dubin1977,Ahlsen1985,Cudeiro2006}), whose functional role may vary between facilitation and suppression of LGN spiking \cite{Cudeiro2006,Alitto2015}. As a result, the algorithm embodied by the retinogeniculate synapse may change if the cortex is active during recordings---however, conducting recordings in the awake state may be technically challenging and the analysis of data from complex models of the retinogeniculate circuit, as for example recently by Rogala and colleagues \cite{Rogala2013}, may pose a viable alternative.

However, investigating information processing at the retinogeniculate synapse while V1 is active would allow us to integrate recordings from V1 as second input to the LGN into our analysis. This approach is an alternative to quantifying prediction errors by measuring information storage in the RGC input alone, thus circumventing the limitations of this approach that were discussed above. We want to highlight that if input from V1 or other sources became available in data sets, the presented framework is easily extendable to include such sources. In this case, one may define the predictive information storage,

\begin{equation}\label{eq:V1sidechannel}
    pAIS = i(RGC^+:RGC^-,V1^-),
\end{equation}

\noindent as information provided by both the past of V1 and the RGC input about the RGC's future state.

Similar to the analysis of local storage-transfer correlations above, the PID-based analysis of coding for predictable versus coding for unpredictable information can be adapted to an additional input: if the synapse coded for prediction errors, we would expect information transfer only in case of a mismatch between top-down input from V1 and bottom-up input from the RGC. In other words, the LGN should spike whenever it received a spike exclusively in the top-down signal \textit{or} in the bottom-up signal. This is measured by the synergistic information, $I_{syn}(LGN^+:RGC^-,V1^-)$. If the synapse coded for predictable input, we would expect information transfer in case of matching inputs, i.e., the LGN should spike whenever it received a spike in \textit{both} input signals. This is measured by the shared information, $I_{shd}(LGN^+:RGC^-,V1^-)$.

\subsection*{Considerations on cortico-cortical predictive coding}

\subsubsection*{Relation of results from the retinogeniculate synapse to theories of predictive coding in the neocortex}

We remark that the results presented here should not be seen as a refutation of predictive coding theories that propose the bottom-up propagation of prediction errors as a general information processing principle in the \emph{neocortex} \cite{Mumford1992,Rao1999,Hohwy2008}. This is simply because we did not investigate information processing in the cortex and the results presented here should thus not be seen as general evidence against this family of theories. Rather, different algorithms may be used at other levels of the processing hierarchy, especially within the cortex \cite{Murray2004,Rathbun2007,Summerfield2008}, even though our study provides evidence for an enhancement of predictable information in the subcortical visual system.

\subsubsection*{Analysing cortico-cortical predictive coding using storage-transfer-correlations}
The analysis presented here relied on the fact that we were able to analyse the transmission of information from the inputs to the output of a single neuron because both the relevant inputs as well as the corresponding outputs were directly recorded. When transferring our analysis framework to cortico-cortical predictive coding, we face two obvious difficulties: first, it will become next to impossible to cover \emph{all} relevant inputs to a neuron with sufficient spatio-temporal resolution (although some in-vivo single-cell optical techniques hold some promise here); second, the estimation of the resulting high-dimensional probability distributions will certainly pose an extreme challenge. At present the best way forward here seems to rely on summary signals such as local field potentials (LFP) or optical techniques with a coarser resolution. This, however, then results in the difficulty that we are no longer in a position to analyse the information transferred through a single neuron. Thus, in this case we will have to resort to analysing a triplet of cortical patches---such that one hierarchically lower patch provides the inputs on which to quantify the $lAIS$, a second patch at an intermediate stage in the hierarchy serves as a receiver, and the information transfer from this second patch to a third one even higher up in the processing hierarchy will provide a measure of information transferred in the outputs of the intermediate, second patch. This idea is presented in more detail in \cite{Wibral2015BitsFromBrains}. Despite these difficulties, the analysis of cortico-cortical predictive coding using the proposed information-theoretic framework seems highly promising, as very explicit predictions on the type of predictive coding, the location of error-computing units in upper cortical layers, and the corresponding LFP-frequency signatures of error signal have been made \cite{Bastos2012,Bastos2015}. Thus, we expect these hypotheses to be directly testable using frequency-resolved measures of information transfer \cite{pinzuti2020measuring}.

\section*{Conclusion}
Tests of predictive coding theories are at risk of being influenced by implicit assumptions of the researchers about what a brain should predict. To circumvent this, careful design of such experiments is necessary, but may not always be possible due to a lack of the required prior knowledge of brain function. Also, such tests cannot not refute of confirm predictive coding theories based on data from other experiments not designed with predictive coding in mind---although such tests need to be performed for a theory that claims a rather broad applicability to brain function. For these reasons an analysis framework that is independent of experimental design and the experimenter's assumptions would be highly beneficial. Here we present such a framework based on the correlation between the information-theoretic equivalents of predictability and prediction errors. In a proof-of-principle analysis of the re-encoding of retinal ganglion cell inputs in the lateral geniculate nucleus principal cells we demonstrate coding for predictable information in an anesthetized animal.

\section*{Supporting information}

\paragraph*{S1 Appendix.}
\label{S1_Appendix}
{\bf Localized bias Panzeri-Treves-correction for plug-in estimators.}

The bias of plug-in entropy-estimators for finite samples sizes can be analytically approximated for asymptotic sampling regimes, i.e., $N \geq |\mathcal{A}_X|$, as shown by Panzeri and Treves \cite{miller1955,panzeri1996correction}:

\begin{align}
    B_{X} &= BIAS \left[ H(X) \right] = \frac{-1}{2N \ln(2)} \left[m - 1\right] \label{eq:pt_h},\\
    B_{X|Y} &= BIAS\left[H(X|Y)\right] = \frac{-1}{2N \ln(2)} \left[\sum_{y \in \mathcal{A}_Y} m_y - 1\right], \label{eq:pt_h_cond}
\end{align}

\noindent where $m$ is the alphabet size $m=|\mathcal{A}_X|$, and $m_y$ is the alphabet size given $Y=y$ has occurred, $m_y=|\mathcal{A}_{X|Y=y}|$.

Note that the true alphabet sizes, $m$ and $m_y$, are typically not known for experimental data and the number of actually observed responses can only be considered a lower bound on the true values. An estimate of $m$ and $m_y$ can be obtained from experimental data via a \textit{Bayesian counting procedure} \cite{panzeri1996correction,panzeri2007bias}, for example implemented in the pyEntropy toolbox \cite{ince2009pyentropy}.

To obtain a local bias correction for $lAIS$ and $lTE$ using the Panzeri-Treves correction, we first provide a bias correction for local MI. We start by applying the correction to a plug-in estimator for the non-local MI (Eq 1, main text),

\begin{equation}
        \hat{I}(X;Y) \approx \sum_{y \in \mathcal{A}_Y}\sum_{x \in \mathcal{A}_X} \left[ \hat{p}(x,y) \log_2 \frac{\hat{p}(x,y)}{\hat{p}(x)p(y)} \right] + B_{X} - B_{X|Y}.
\end{equation}

\noindent To obtain a localized version of this estimator, note that the correction term $B_X$ is constant over $x \in \mathcal{A}_X$ and $y \in \mathcal{A}_Y$, but $B_{X|Y}$ is not. The latter can be rewritten as a sum over $y \in \mathcal{A}_Y$, $B_{X|Y} = \sum_{y \in \mathcal{A}_Y} b_{X|y} = \sum_{y \in \mathcal{A}_Y} \frac{-1}{2N \ln(2)} \left[m_y - 1\right]$, (Eq. \ref{eq:pt_h_cond}), such that we can write 

\begin{equation}
        I(X;Y) \approx \sum_{y \in \mathcal{A}_Y} \left[\sum_{x \in \mathcal{A}_X} \left[ \hat{p}(x,y) \log_2 \frac{\hat{p}(x,y)}{\hat{p}(x)p(y)} \right] - b_{X|y} \right] + B_{X},
\end{equation}

\noindent where $b_{X|y}$ is the individual contribution of realization $y \in \mathcal{A}_Y$ to the average correction term $B_{X|Y}$.
By dividing by the alphabet size $m$, we can rewrite this as

\begin{equation}
        I(X;Y) \approx \sum_{y \in \mathcal{A}_Y} \left[\sum_{x \in \mathcal{A}_X} \left[ \hat{p}(x,y) \log_2 \frac{\hat{p}(x,y)}{\hat{p}(x)p(y)} - \frac{b_{X|y}}{m} \right] \right] + B_{X}.
\end{equation}

We may now rewrite the corrected MI as the expected value over all observations, analogous to \cite{lizier2013book}. The constant term $B_X$ can be brought inside the average because of the linearity of the expected value, 

\begin{equation}
    \begin{aligned}
        I(X;Y) &\approx \sum_{y \in \mathcal{A}_Y, x \in \mathcal{A}_X} \left[ \hat{p}(x,y) \log_2 \frac{\hat{p}(x,y)}{\hat{p}(x)p(y)} - \frac{b_{X|y}}{m}\right] + B_{X}\\
        &= \sum_{y \in \mathcal{A}_Y, x \in \mathcal{A}_X} \left[ \frac{1}{N}\sum_{g=1}^{c(x,y)}\log_2 \frac{\hat{p}(x,y)}{\hat{p}(x)p(y)} - \frac{b_{X|y}}{m}\right] + B_{X}\\
        &= \frac{1}{N}\sum_{i=1}^N \left[\log_2 \frac{\hat{p}(x,y)}{\hat{p}(x)p(y)} - \frac{b_{X|y}}{m} \right] + B_{X}\\
        &= \Bigl\langle i(x_n;y_n) - \frac{b_{X|y_n}}{m} + B_{X}\Bigl\rangle_n \, .\\
    \label{eq:pt_local}
    \end{aligned}
\end{equation}

From this expression, we can immediately formulate the PT-corrected estimators for $lAIS$,

\begin{equation}
    \begin{aligned}
        AIS(X, n, j) &= \biggl\langle lAIS^{PT}(X, n) \biggl\rangle_n \\
        &= \biggl\langle i(\mathbf{x}_{n-1};x_n) + B_{\mathbf{X}_{n-1}} - \frac{b_{\mathbf{X}_{n-1}|x_n}}{|\mathcal{A}_{\mathbf{X}_{n-1}}|} \biggl\rangle_n\, ,\\
        \label{eq:lais_corrected}
    \end{aligned}
\end{equation}

\noindent and for $lTE$ (again because of the linearity of the expected value),

\begin{equation}
    \begin{aligned}
        TE_{SPO}(X& \rightarrow Y, u) = \biggl\langle lTE_{SPO}^{PT}(X \rightarrow Y, u) \biggl\rangle_n \\
        &= I \left( Y_n;\mathbf{X}_{n-u}| \mathbf{Y}_{n-1} \right) \\[0.5em]
        &= \hat{I} \left(\mathbf{X}_{n-u}; Y_n, \mathbf{Y}_{n-1} \right) + B_{\mathbf{X}_{n-u}} - B_{\mathbf{X}_{n-u}| Y_n, \mathbf{Y}_{n-1}} \\
        &\hphantom{XX} - \hat{I} \left( \mathbf{X}_{n-u}; \mathbf{Y}_{n-1} \right) - B_{\mathbf{X}_{n-u}} + B_{\mathbf{X}_{n-u}| \mathbf{Y}_{n-1}} \\[1em]
        &= \biggl\langle i(\mathbf{x}_{n-u}; y_n, \mathbf{y}_{n-1}) + B_{\mathbf{X}_{n-u}} - \frac{b_{\mathbf{X}_{n-1}|y_n, \mathbf{y}_{n-1}}} {|\mathcal{A}_{\mathbf{X}_{n-u}}|}\\
        &\hphantom{XX}- i(\mathbf{x}_{n-u}; \mathbf{y}_{n-1}) + B_{\mathbf{X}_{n-u}} - \frac{b_{\mathbf{X}_{n-1}| \mathbf{y}_{n-1}}}{|\mathcal{A}_{\mathbf{X}_{n-u}}|} \biggl\rangle_n\, .\\
        \label{eq:lte_corrected}
    \end{aligned}
\end{equation}

\paragraph*{S1 Table.}
\label{S1_Table}
{\bf Spike train statistics.} Contribution and efficacy values are taken from \cite{Rathbun2010}.
\begin{table}[H]
	\centering

	\begin{tabular}{lrcrcrccc} \hline
		Pair & \multicolumn{1}{c}{N} & delay & \multicolumn{2}{c}{RGC} & \multicolumn{2}{c}{LGN} & efficacy\textsuperscript{a} & contribution\textsuperscript{b} \\
		   &           &   & no. spikes &  rate & no. spikes &  rate &      &     \\ \hline
		1  &   710,597 & 3 &     20,418 & 0.029 &      7,357 & 0.010 &  5.6 & 15.4 \\
		2  &   709,555 & 3 &     40,316 & 0.057 &      3,849 & 0.005 &  4.1 & 42.5 \\
		3  &   710,302 & 2 &     31,064 & 0.044 &      8,179 & 0.012 & 13.8 & 52.6 \\
		4  &   709,019 & 2 &     20,267 & 0.029 &      9,830 & 0.014 & 15.9 & 32.8 \\
		5  & 2,374,622 & 2 &     86,000 & 0.036 &     22,212 & 0.009 &  1.7 & 6.4 \\
		6  &   471,740 & 2 &     18,200 & 0.039 &      3,055 & 0.006 &  1.7 & 10.3 \\
		7  &   709,014 & 3 &      9,925 & 0.014 &      6,410 & 0.009 &  2.9 & 4.5 \\
		8  &   710,181 & 3 &     29,057 & 0.041 &      9,239 & 0.013 &  3.6 & 11.4 \\
		9  &   710,861 & 3 &     15,800 & 0.022 &      4,126 & 0.006 &  1.7 & 6.6 \\
		10 &   710,892 & 3 &     39,169 & 0.055 &      4,788 & 0.007 &  9.3 & 75.9 \\
		11 &   710,892 & 3 &     39,169 & 0.055 &      3,174 & 0.004 &  5.0 & 61.6 \\
		12 & 1,186,696 & 6 &     11,816 & 0.010 &     22,141 & 0.019 &  4.8 & 2.6 \\
		13 &   709,706 & 2 &     23,048 & 0.032 &     20,982 & 0.030 & 34.6 & 38.0 \\
		14 &   374,866 & 2 &     25,139 & 0.067 &      6,593 & 0.018 &  4.9 & 24.6 \\
		15 &   710,723 & 2 &      8,979 & 0.013 &      8,858 & 0.012 &  2.4 & 2.4 \\
		16 &   710,823 & 2 &     15,901 & 0.022 &     30,151 & 0.042 & 13.0 & 6.8 \\
		17 &   471,876 & 4 &      4,597 & 0.010 &      2,139 & 0.005 & 10.8 & 25.2 \\
		\hline
	\end{tabular}
	\begin{flushleft} \textsuperscript{a} percentage of RGC spikes preceding an LGN spike. \textsuperscript{b} percentage of LGN spikes preceded by RGC spikes.
	\end{flushleft}

\end{table}

\paragraph*{S2 Table.}
\label{S2_Table}
{\bf Optimized embedding lengths and information transfer delays.}
Optimized non-uniform embedding lengths for $lAIS$ and $lTE$ estimation and reconstructed information-transfer delay $u$ for $lTE$ estimation.

\begin{table}[H]
	\centering
    \begin{tabular}{@{}lllllllllllllllll@{}} \hline 
        Cell Pair	 & 1 & 2 & 3 & 4 & 6 & 7 & 8 & 9 & 10 & 11 & 12 & 13 & 14 & 15 & 16 & 17\\ \hline
        $lAIS$       & 8 & 9& 9& 9& 7& 8& 9& 7& 10& 10& 6& 8& 7& 4& 7& 4 \\
        $lTE$ source & 4 & 4& 4& 2& 3& 1& 2& 3& 4& 5& 1& 2& 2& 1& 3& 3 \\
        $lTE$ target & 5 & 6& 7& 9& 3& 7& 7& 4& 7& 5& 8& 7& 6& 6& 5& 5 \\
        delay $u$    & 3 & 3& 2& 2& 2& 3& 3& 3& 3& 3& 6& 2& 2& 2& 2& 4 \\
        \hline
    \end{tabular}
\end{table}

\paragraph*{S3 Table.}
\label{S3_Table}
{\bf Local storage-transfer correlation coefficients.}
Local storage-transfer correlation (LSTC) coefficients for all cell pairs with significant $lAIS$ and $lTE$.

\begin{table}[H]
	\centering
    \begin{tabular}{@{}llll@{}}
    \hline

    Cell Pair & $c(LAIS, LTE)$ & p-value & contribution $[\%]$\\ \hline
        1	& 0.0454   & 0.0000  & 15.40 \\
        2	& 0.0485   & 0.0000  & 42.50 \\
        3	& 0.1821   & 0.0000  & 52.60 \\
        4	& 0.1878   & 0.0000  & 32.80 \\
        6	& 0.0149   & 0.0000  & 10.30 \\
        7	& 0.0285   & 0.0000  & 4.50 \\
        8	& 0.0536   & 0.0000  & 11.40 \\
        9	& 0.0331   & 0.0000  & 6.60 \\
        10	& 0.2675   & 0.0000  & 75.90 \\
        11	& 0.1884   & 0.0000  & 61.60 \\
        12	& 0.0695   & 0.0000  & 2.60 \\
        13	& 0.0443   & 0.0000  & 38.00 \\
        14	& -0.0209  & 1.0000  & 24.60 \\
        15	& 0.0056   & 0.0040  & 2.40 \\
        16	& 0.1866   & 0.0000  & 6.80 \\
        17	& 0.0827   & 0.0000  & 25.20 \\
    \hline
    \multicolumn{4}{l}{$c(LSTC, contribution)=0.6879$, $p = 0.0030$}
    \end{tabular}
\end{table}

\paragraph*{S1 Fig.}
\label{S1_Fig}
{\bf Distribution of optimized past variable lags.} Distribution of lags of past variables identified through optimization of the non-uniform embedding for (A) $lAIS$, (B) $lTE$ (source), and (C) $lTE$ target.

\begin{figure}[H]
	\includegraphics{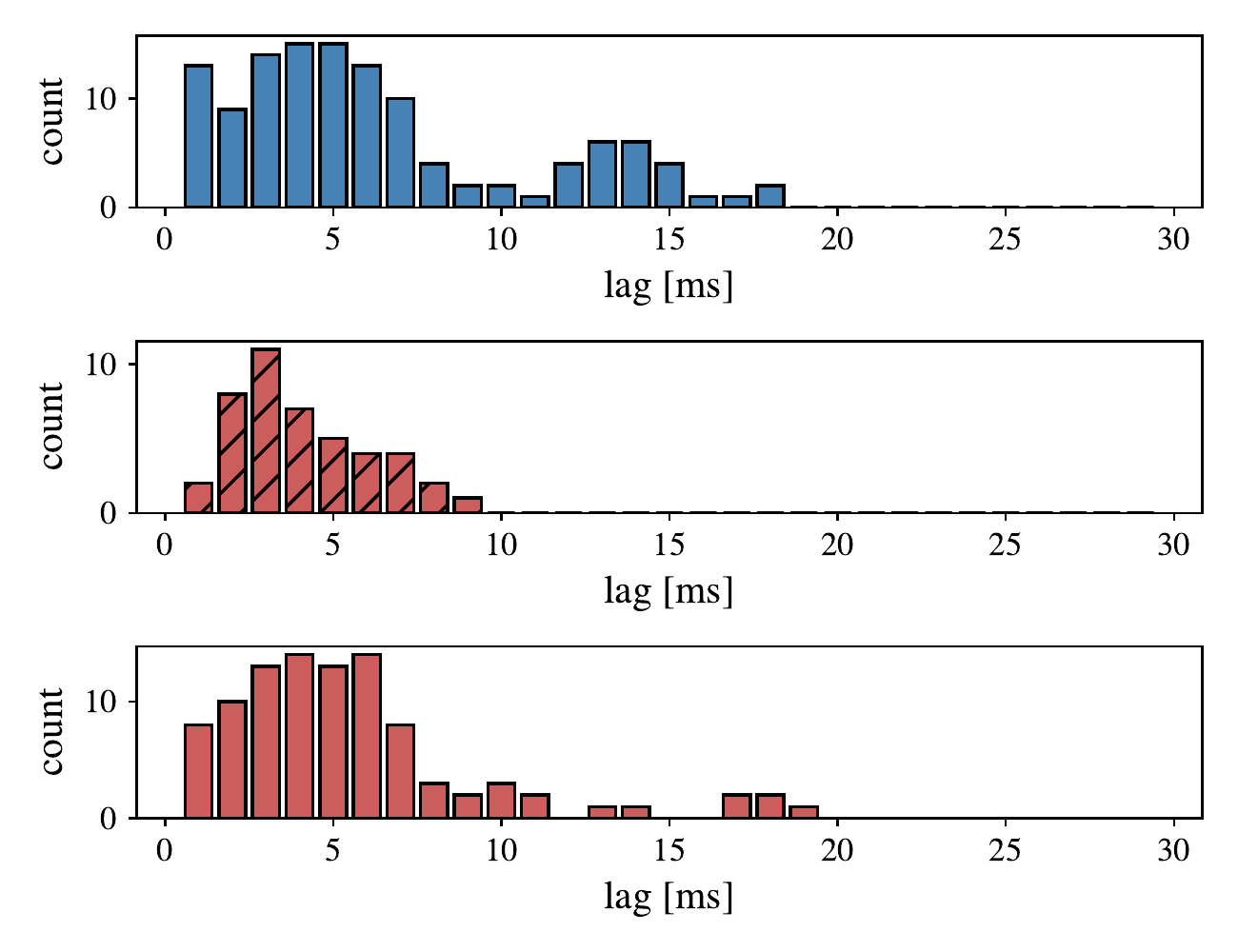}
\end{figure}

\paragraph*{S2 Fig.}
\label{S2_Fig}
{\bf Classification of relayed versus non-relayed RGC spikes.} Classification accuracy of relayed versus non-relayed spikes from $lAIS$ values, inter-spike intervals (ISI), and spike counts within a time window of \SI{30}{\milli\second} prior to an RGC spike (k-nearest neighbor classifier with $k = 5$, average values over ten repetitions of five-fold cross validation, error bars indicate one standard deviation). The dotted line indicates the performance of a baseline classifier, predicting relayed spikes with probability corresponding to the relative frequency of relayed spikes in the training set.

\begin{figure}[H]
	\includegraphics{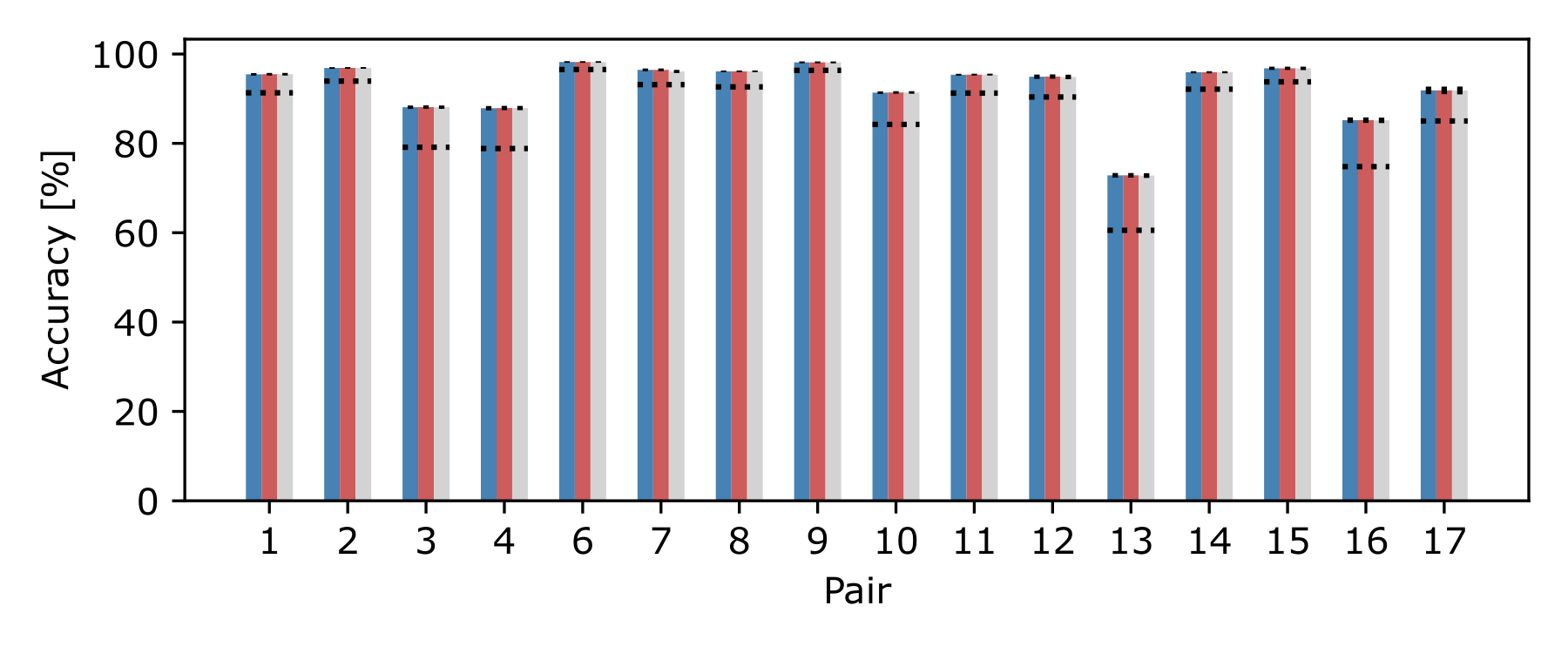}
	\label{fig:s2}
\end{figure}

\section*{Acknowledgments}
The authors thank Prescott C. Alexander of the University of California, Davis, for setting up the data repository and his support in using the data provided there.

\nolinenumbers


\end{document}